\documentclass[sigconf,nonacm,screen=true]{acmart}

\usepackage{listings}
\usepackage{subcaption}

\let\oldFootnote\footnote
\newcommand\nextToken\relax

\renewcommand\footnote[1]{%
    \oldFootnote{#1}\futurelet\nextToken\isFootnote}

\newcommand\isFootnote{%
    \ifx\footnote\nextToken\textsuperscript{,}\fi}

\begin{document}
\title{The gem5 Simulator: Version 20.0+}
\subtitle{A new era for the open-source computer architecture simulator}
\author{
  Jason Lowe-Power, Abdul Mutaal Ahmad, Ayaz Akram, Mohammad Alian, Rico Amslinger, Matteo Andreozzi, Adri\`a Armejach, Nils Asmussen, Brad Beckmann, Srikant Bharadwaj, Gabe Black, Gedare Bloom, Bobby R. Bruce, Daniel Rodrigues Carvalho, Jeronimo Castrillon, Lizhong Chen, Nicolas Derumigny, Stephan Diestelhorst, Wendy Elsasser, Carlos Escuin, Marjan Fariborz, Amin Farmahini-Farahani, Pouya Fotouhi, Ryan Gambord, Jayneel Gandhi, Dibakar Gope, Thomas Grass, Anthony Gutierrez, Bagus Hanindhito, Andreas Hansson, Swapnil Haria, Austin Harris, Timothy Hayes, Adrian Herrera, Matthew Horsnell, Syed Ali Raza Jafri, Radhika Jagtap, Hanhwi Jang, Reiley Jeyapaul, Timothy M. Jones, Matthias Jung, Subash Kannoth, Hamidreza Khaleghzadeh, Yuetsu Kodama, Tushar Krishna, Tommaso Marinelli, Christian Menard, Andrea Mondelli, Miquel Moreto, Tiago M\"uck, Omar Naji, Krishnendra Nathella, Hoa Nguyen, Nikos Nikoleris, Lena E. Olson, Marc Orr, Binh Pham, Pablo Prieto, Trivikram Reddy, Alec Roelke, Mahyar Samani, Andreas Sandberg, Javier Setoain, Boris Shingarov, Matthew D. Sinclair, Tuan Ta, Rahul Thakur, Giacomo Travaglini, Michael Upton, Nilay Vaish, Ilias Vougioukas, William Wang, Zhengrong Wang, Norbert Wehn, Christian Weis, David A. Wood, Hongil Yoon, \'Eder F. Zulian
}
\authornote{Email Jason Lowe-Power (jlowepower@ucdavis.edu) with all questions and comments.}

\titlenote{
    gem5 is the result of the merger of the GEMS project started in 1999, and the m5 project started in 2003.
    Development of gem5 has been active for about 20 years, and this version is being published in 2020. Thus, ``gem5-20''.
}

\begin{abstract}
    The open-source and community-supported gem5 simulator is one of the most popular tools for computer architecture research.
    This simulation infrastructure allows researchers to model modern computer hardware at the cycle level, and it has enough fidelity to boot unmodified Linux-based operating systems and run full applications for multiple architectures including x86, Arm\textregistered, and RISC-V.
    The gem5 simulator has been under active development over the last nine years since the original gem5 release.
    In this time, there have been over 7000 commits to the codebase from over 250 unique contributors which have improved the simulator by adding new features, fixing bugs, and increasing the code quality.
    In this paper, we give an overview of gem5's usage and features, describe the current state of the gem5 simulator, and enumerate the major changes since the initial release of gem5.
    We also discuss how the gem5 simulator has transitioned to a formal governance model to enable continued improvement and community support for the next 20 years of computer architecture research.
\end{abstract}

\maketitle
\renewcommand{\shortauthors}{Lowe-Power and the gem5 Community}

\section{The gem5 Simulator}


There is ``a new golden age for computer architecture''~\cite{HennessyPatterson-turingLect-isca18, HennessyPatterson-CACM19} driven by changes in technology (e.g., the slowdown of Moore's Law and Dennard Scaling) and ever increasing computational needs.
One of the first steps in research and development of new hardware architectures is software-based modeling and simulation.
The gem5 simulator~\cite{Binkert-gem5-2011} is currently one of the most popular academic-focused computer architecture simulation frameworks.
Since its publication in 2011, the gem5 paper has been cited over 3600 times\footnote{\url{https://scholar.google.com/scholar?q=gem5}}, and every year many papers published in the top computer architecture venues use gem5 as their main evaluation infrastructure.
Additionally, gem5 is one of the tools used to design the Fugaku supercomputer, one of the first exascale systems~\cite{Kodama:Riken:2019}.

The gem5 simulator~\cite{Binkert-gem5-2011} is an open source community-supported computer architecture simulator system.
It consists of a simulator core and parametrized models for a wide number of components from out-of-order processors, to DRAM, to network devices.
The gem5 project consists of the gem5 simulator\footnote{\url{https://gem5.googlesource.com/public/gem5}}, documentation\footnote{\url{https://www.gem5.org/}}, and common resources\footnote{\url{https://gem5.googlesource.com/public/gem5-resources}} that enable computer architecture research.

The gem5 project is governed by a meritocratic, consensus-based community governance document\footnote{\url{https://www.gem5.org/governance/}} with a goal to provide a tool to further the state of the art in computer architecture.
The gem5 simulator can be used for (but is not limited to) computer-architecture research, advanced development, system-level performance analysis and design-space exploration, hardware-software co-design, and low-level software performance analysis.
Another goal of gem5 is to be a common framework for computer architecture research.
A common framework in the academic community makes it easier for other researchers to share workloads and models as well as compare and contrast their innovations with other architectural techniques.

The gem5 community strives to balance the needs of its three categories of users: academic researchers, industry researchers, and students learning computer architecture.
For instance, the gem5 community strives to balance adding new features (important to researchers) and a stable code base (important for students).
Specific user needs important to the community are enumerated below:
\begin{itemize}
    \item Effectively and efficiently emulate the behavior of modern processors in a way that balances simulation performance and accuracy
    \item Serve as a malleable baseline infrastructure that can easily be adapted to emulate the desired behaviors
    \item Provide a core set of APIs and features that remain relatively stable
    \item Incorporate features that make it easy for companies and research groups to stay up to date with new features and bug fixes as well as continue contributing to the project
    \item Additionally, the gem5 community is committed to openness, transparency, and inclusiveness.
\end{itemize}

In this paper, we discuss the current state of gem5.
We first discuss the past, present and future of the gem5 project and how to become a member of the gem5 community for researchers, students, and teachers in Section~\ref{sec:current-gem5}.
Then, we give an overview of gem5's main features available today and describe how to use gem5 for its main use case: computer architecture simulation~\ref{sec:main-features}.
Finally, Section~\ref{sec:changes} enumerates the major changes in gem5 in the past nine years since the initial release.

It has taken a huge number of people to make gem5 what it is today.
One of the goals of this paper is to recognize the hard work on this community infrastructure which is often overlooked.
To that end, we strive to include everyone who contributed to gem5 and document as many of the major changes as we can.
Section~\ref{sec:acks} acknowledges those contributors who are not authors of this paper and without whom gem5 would not be as successful.

\subsection{The Past, Present, and Future of gem5}\label{sec:current-gem5}

The gem5 simulator was born when the m5 simulator~\cite{BinkertDHLSR06} created at University of Michigan merged with the GEMS simulator~\cite{MartinSBMXAMHW05} from University of Wisconsin.
These were two academic-oriented simulators, neither of which had an open development community (both simulators had their source available for free\footnote{\url{https://sourceforge.net/projects/m5sim/}}\footnote{\url{https://research.cs.wisc.edu/gems/home.html}}, but did not have a community-oriented development process).
Both of these simulators were quite popular on their own.
The GEMS paper has been cited over 1800 times and the m5 paper has been cited over 1000 times.

\begin{figure}
    \centering
    \begin{subfigure}{0.9\linewidth}
      \centering
      \includegraphics[width=\linewidth]{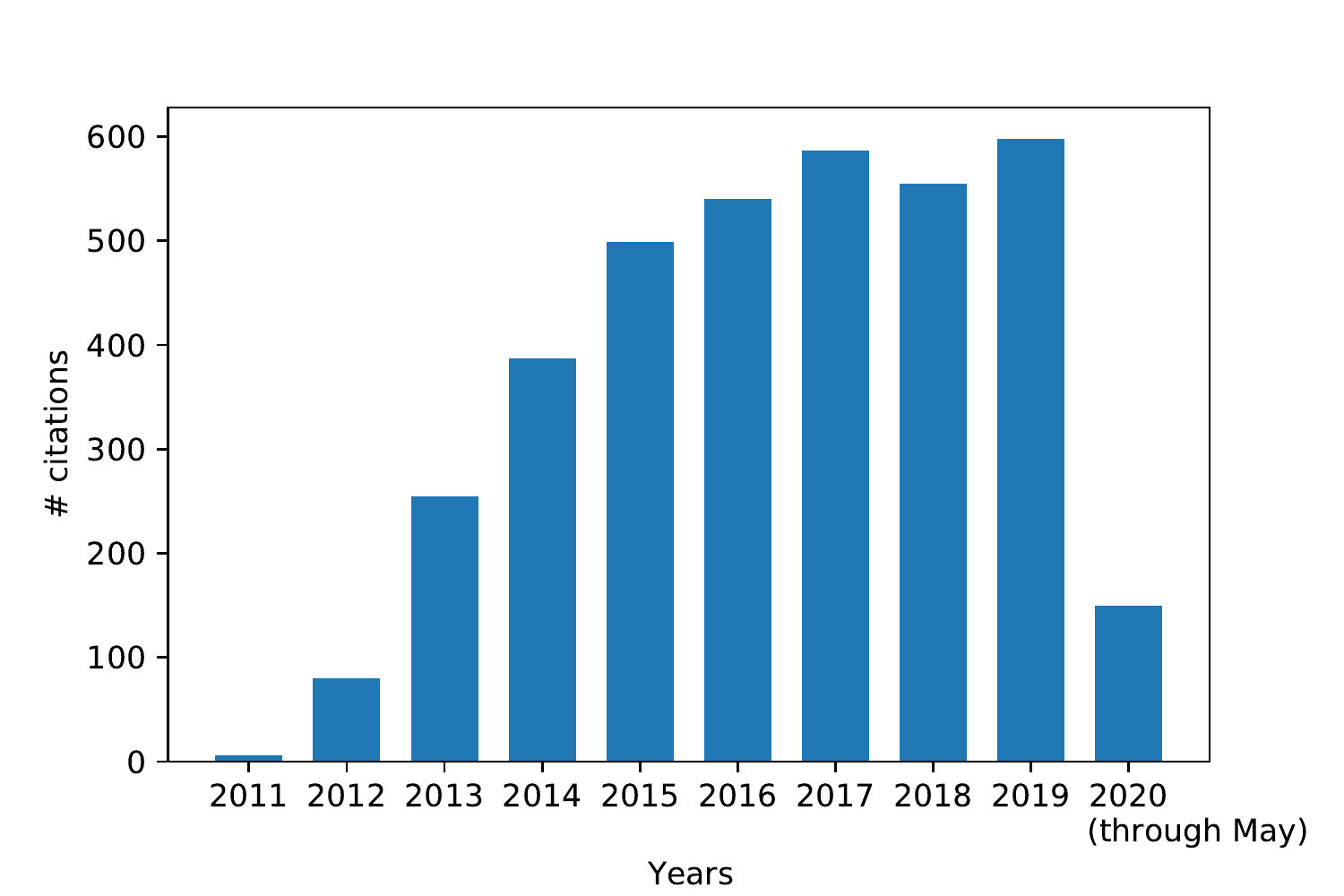}
      \caption{Number of citations}
      \label{fig:citations}
    \end{subfigure}

    \begin{subfigure}{0.9\linewidth}
      \centering
      \includegraphics[width=\linewidth]{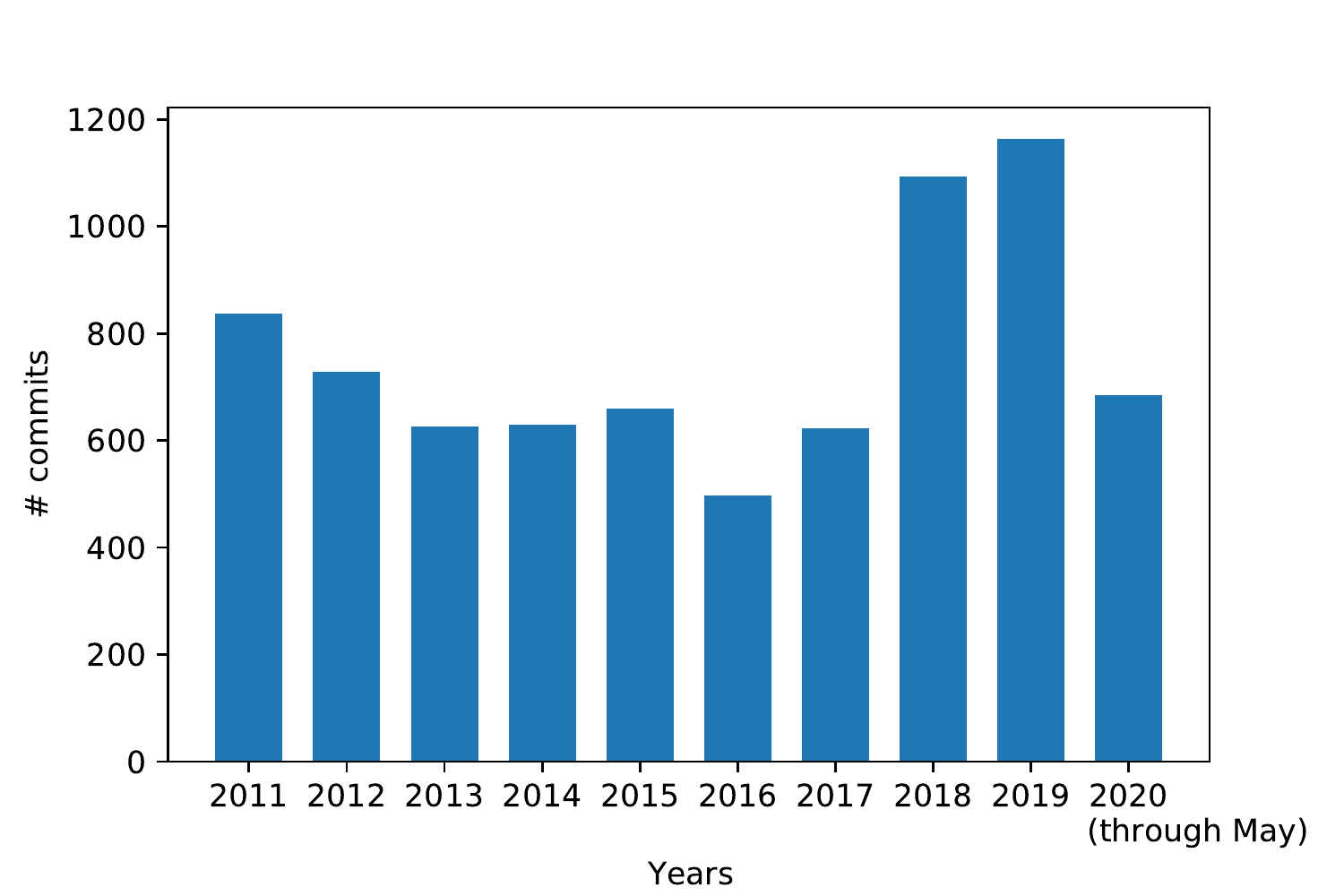}
      \caption{Number of commits}
      \label{fig:commits}
    \end{subfigure}

    \begin{subfigure}{0.9\linewidth}
      \centering
      \includegraphics[width=\linewidth]{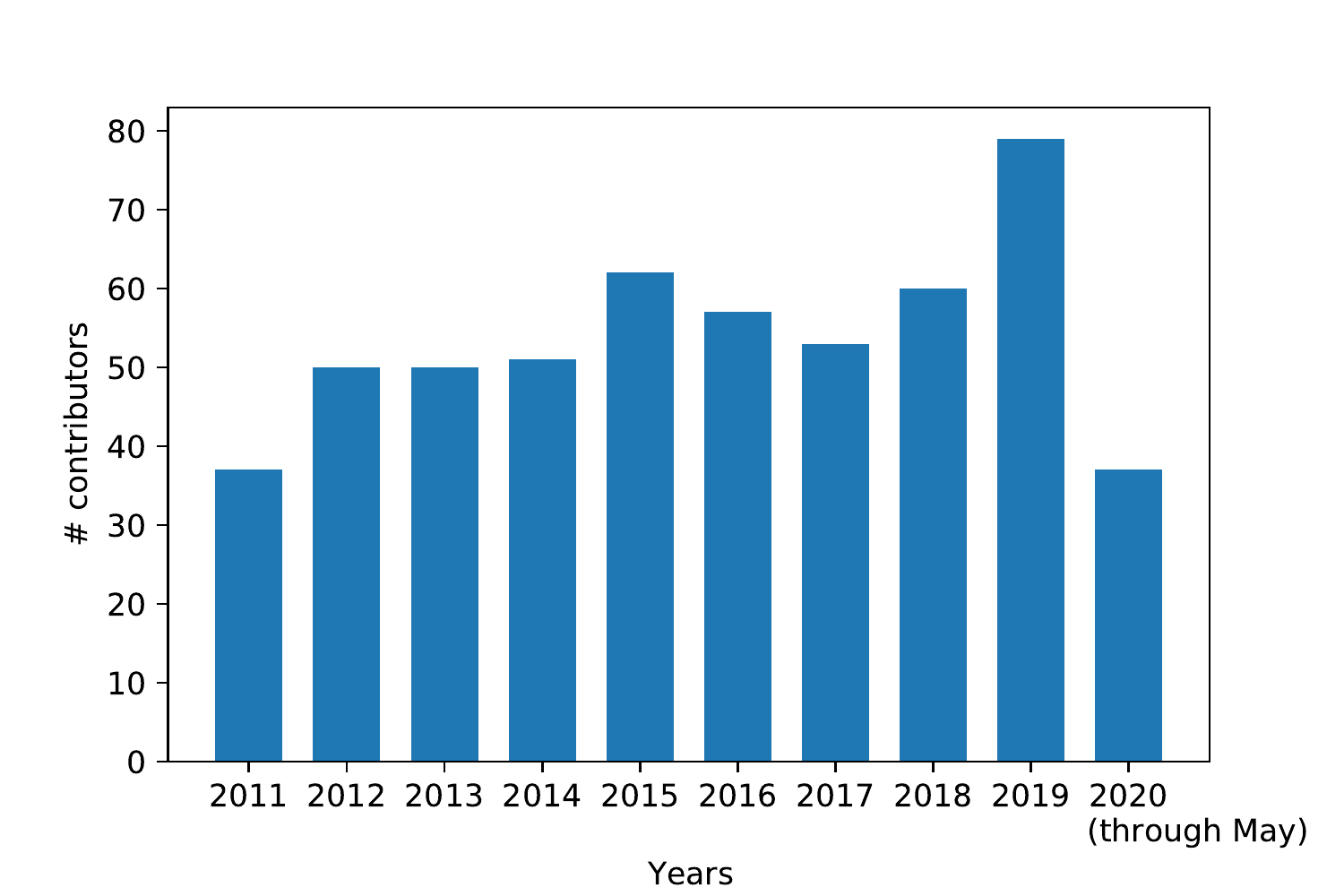}
      \caption{Number of contributors}
      \label{fig:contributors}
    \end{subfigure}
    \caption{Number of gem5 citations, commits and contributors from 2011 to May 2020.}
    \label{fig:gem5_citations_commits_contributors}
\end{figure}

Since its initial release nine years ago the gem5 simulator has been wildly successful.
In this time, the use of gem5 has exploded.
Although not a perfect metric, as shown in Figure~\ref{fig:citations} the gem5 paper has received over 3600 citations according to Google Scholar.

At the same time, the contributor community has also grown.
Figure~\ref{fig:commits} shows the number of commits per year and Figure~\ref{fig:contributors} shows the number of unique contributors per year.
These figures show that since the initial release of gem5 in 2011, development has been accelerating.

With this acceleration in use and development came growing pains~\cite{Power-gem5horrors-2015}.
The gem5 community was going through a shift, from a small project with most contributors from one or two academic labs, to a project with worldwide-distribution of contributors.
Additionally, given the growing user base, we could no longer assume that all gem5 users were also going to be main developers.

To solve the problems brought up by the expanding gem5 community, the gem5 project has made major changes in the past nine years.
We now have a formal governance structure, we have improved our documentation (see Section~\ref{sec:learning}), we have moved to a better distributed development platform, and we have improved our community outreach.

To institute a formal governance model, we followed the best practices from other successful open source projects.
We chose to institute a meritocratic governance model where anyone with an interest in the project can join the community, contribute to the project design and participate in the decision-making process.
The governance structure also defines the roles and responsibilities of members of the community including users, contributors, and maintainers.
We also formed a project management committee~(PMC) with a mix of industry and academic members to help ensure smooth running of the project.

To simplify the contribution process, we have instituted many industry-standard development methodologies including providing a \lstinline|CONTRIBUTING| document in the gem5 source.
In the past, gem5 code contributions were managed with a number of esoteric software packages.
Now, all gem5 code is stored in a git repository\footnote{\url{https://gem5.googlesource.com/}}, code review is managed on gerrit\footnote{\url{https://gem5-review.googlesource.com/}}, we have continuous integration support (see Section~\ref{sec:testing}), our website is implemented with Jekyll and markdown\footnote{\url{https://gem5.googlesource.com/public/gem5-website}}, and we have a Jira-based issue tracker\footnote{\url{https://gem5.atlassian.net/}}.

After transitioning to these more well known tools and improving our development practices, we have seen a further rise in the number of community contributors and using gem5 has become easier.
Continuous integration enables us to test every single changeset before it is committed.
This allows us to catch bugs \emph{before} they are committed into the mainline repository which makes gem5 more stable.
Similarly, by implementing a bug tracking system, we can track issues that affect gem5.
For example, in the first six months of using a bug tracker we have closed over 250 issues.

\subsubsection*{The future of gem5}
The future of gem5 is bright.
We are continuing to work with the community to define the roadmap for gem5 development for the next version of gem5, version 20.1, and beyond.
In the short term, we are excited about improvements to the underlying infrastructure of gem5 with better testing, refactoring of aging code (some of gem5's code is over 20 years old!), and adding well-defined stable APIs.
By defining stable APIs, we will make it easier for the community to build off of gem5.
For instance, the inter-simulator interface is currently being defined so that gem5 can be used in conjunction with other simulators (e.g., SST~\cite{RodriguesHemmert2011-sst, HsiehPedretti2012-sst-gem5}, SystemC (Section ~\ref{sec:systemc}), and many others).
We are also working on improving the interconnect model (Section~\ref{sec:garnet}), adding support for non-volatile memory controllers (Section~\ref{sec:nvm}), and a graphical user interface (GUI).

One of the most exciting features coming to gem5 is that we will provide the community with a set of \emph{publicly validated} models and parameters which will model current architectural system components including CPU cores, GPU compute units (CUs), caches, main memory systems, and devices.
Past research~\cite{butko2012accuracy, nowatzki2015architectural, endo2014micro, akram201686, asri2016simulator, akram2019validation, gutierrez2014sources, jo2018diagsim, tanimoto2017dependence, walker2018hardware} has shown that some gem5 models can be imprecise.
We strive for accuracy compared to real systems; however, since most systems are proprietary and complex, accuracy for all workloads will be difficult.
Thus, we will broadly advertise the relative performance, power, and other metrics when providing these models so users can make an informed decision when choosing their baseline configurations.
This will reduce the researcher's time spent on configuring baselines and allow them to concentrate more effort on analyzing and developing their novel research ideas.
The first step towards this goal of validated baselines is the gem5 resources repository described in Section~\ref{sec:resources}.

Finally, we are planning to publish an online \emph{Learning gem5} course based on an expanded version of the \emph{Learning gem5} material~\ref{sec:learning}\footnote{\url{http://www.gem5.org/documentation/learning_gem5/}}.
This course will cover how to get started using gem5, how to develop new models to be used in gem5, and the details of gem5's software architecture.
In addition to the online version of the course, we will continue to conduct tutorials and workshops at computer architecture and computer systems conferences.

However, the broader gem5 community is the most important part of gem5's future.
In the next section, we discuss how to become part of the gem5 community.

\subsection{Becoming Part of the gem5 Community}

As a reader of this paper, you are already becoming part of the gem5 community!
Anyone who uses gem5 or contributes in any way is part of the gem5 community.
Contributing can be as simple as sending a question on the gem5 mailing list\footnote{\url{http://www.gem5.org/mailing_lists/}} or as complex as adding a new model to the upstream codebase.
Below, we discuss some of the common ways to use gem5 and become part of the community.

\subsubsection{For Researchers}

Currently, the most common gem5 use case is computer architecture research.
In this case, researchers download the gem5 sources, build the simulator, and then add their own device models on top of the models included in upstream gem5.
This use case requires deep knowledge of the core simulation frameworks of gem5.
However, we are working to make it easier to get started developing and researching with gem5 through efforts like the \emph{Learning gem5} materials and online course (Section~\ref{sec:learning}).

After using gem5 in their research, we encourage these users to contribute their improvements and fixes to gem5 back to the mainline codebase.
Not only does this improve gem5 for others, but it also makes reproducing research results easier.
Rather than managing many local changes and trying to keep up with new releases of gem5, when code is contributed upstream it is the responsibility of \emph{others in the community} to ensure that the code stays up to date.
Additionally the gem5 project employs a permissive BSD license to lower the barrier of contribution for both academic and industry researchers.

\subsubsection{For Students and Teachers}

The gem5 simulator can be used as a tool for teaching computer architecture as well.
Historically, there has been a very steep learning curve for using gem5 even for simple experiments.
However, we are improving the documentation for new users.

We will be continuing to improve gem5 with the goal of making it easier for both students and teachers to learn and teach computer architectures concepts.
For example, the new \emph{Learning gem5} material created for the online course will include a set of example exercises that we hope can be used in both undergraduate and graduate computer architecture courses.
Additionally, we are working to develop a new graphical front-end for gem5 and to develop known-good models that do not required deep knowledge of simulator internals to configure and use.

gem5 can be used in a computer architecture course by having the students download and build gem5 themselves or by providing them with a pre-built binary.
Then, the students can create different gem5 configurations which vary hardware parameters (e.g., issue width, cache associativity, etc.).
Finally, the students can explore the effects of these architectural changes on a wide array of common benchmarks and realistic applications from the gem5-resources repository (see Section~\ref{sec:resources}).

\subsection{gem5's Main Features}
\label{sec:main-features}

The gem5 simulator is mainly used to conduct computer architecture research.
In most cases, researchers have an application or benchmark for which they want to measure some \emph{statistics} under different hardware configurations.
For instance, they may be interested in the run time, the memory bandwidth, number of branch predictor mis-speculations, etc.
The gem5 simulator allows users to run applications and simulate the timing of hardware structures.
It contains parameterized models for the processor core (CPU), memory, and devices.

Although there are many computer architecture simulators, and many of these are open source with features that overlap with gem5, gem5 is a unique simulation infrastructure.
\begin{itemize}
    \item gem5 is \emph{dynamically configurable} through a robust Python-based scripting interface. Most other simulators are configured statically with flat text files (e.g., json) or at compilation time. On the other hand, gem5 allows users to simulate complex systems much more easily by using object-oriented Python scripts to compose simpler systems into more complex ones.
    \item gem5 is \emph{extensible} through a clean model API. The gem5 simulator has over 300 parameterized models and adding new models and parameters is straightforward and well documented.
    \item gem5 is a \emph{full system} simulator. Its high-fidelity models can support booting unmodified operating systems and running unmodified applications with cycle-level statistics.
    \item gem5 is a \emph{community-driven} and \emph{frequently updated} project. The gem5 community is thriving. Since its original release nine years ago, there have been over 250 unique contributors and over 7000 commits. Even in the last six months, gem5 has had over 850 commits and 50 unique contributors.
\end{itemize}

\begin{figure*}
  \centering
  \includegraphics[width=0.7\linewidth]{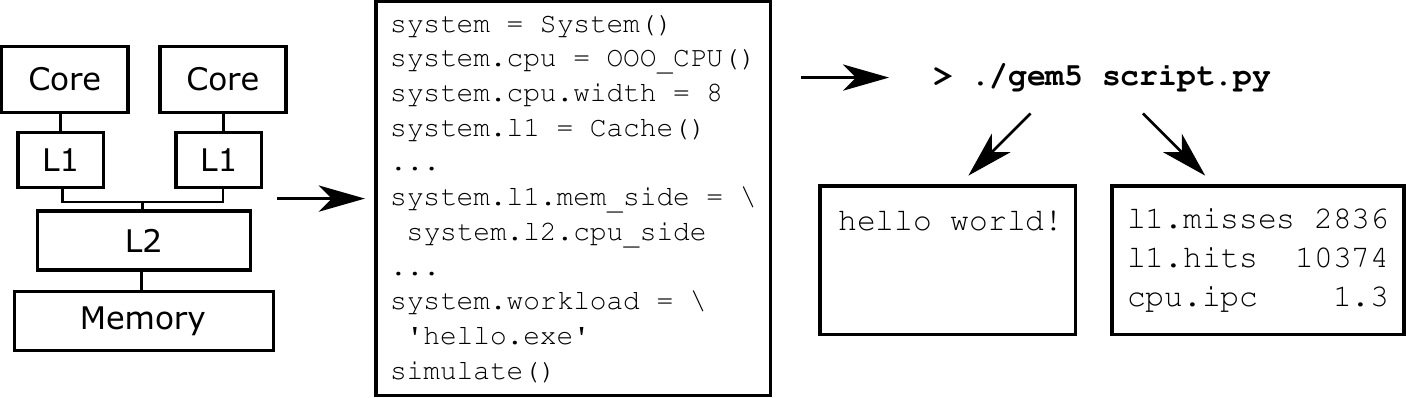}
  \caption{How gem5 is used for computer architecture research. An example system is shown on the left, the sketch of a simulation script is shown in the middle, and the results of the gem5 simulation are shown on the right.}
  \label{fig:gem5-usage}
\end{figure*}

An overview of gem5's usage is shown in Figure~\ref{fig:gem5-usage}.
First, the user chooses a system to simulate (e.g., a two core system with two levels of cache shown on the left of Figure~\ref{fig:gem5-usage}).
Then, the user writes a Python script that describes the system under test by instantiating model objects (\lstinline|SimObjects| in gem5 terminology).
Each object has a number of parameters that the user can modify in their script by setting member variables of the Python objects.
This script is also used to control the simulator and can start the simulation, stop the simulation, save the simulation state, and complete other simulator interactions.
To execute the simulation, the user passes this Python script to the gem5 binary, which acts as a Python interpreter and executes the script.
This instantiates the system and runs the simulation as specified in the script.
The output of gem5 is the output of the application (e.g., standard output or the serial terminal in full system mode) and statistics for each of the simulated model objects.

Many of gem5's features are also useful for other computer systems and programming languages research (e.g., as a platform for developing JIT compilers~\cite{Shingarov2015-jit}).

The gem5 simulator is an ``execute-in-execute'' simulator.
Each operation (e.g., instruction, memory request, I/O operation) is functionally completed at the point that the timing simulation specifies.
For instance, each instruction is executed when it is in the execute stage of the pipeline.
This is in contrast to trace-based and execute-ahead simulators.
The main benefit of execute-in-execute is when running applications whose code path depends on the timing of multiple threads or I/O.
Trace-based or execute-ahead simulation may hide potential behaviors and produce different timing results than real hardware.

To enable modularity, gem5 separates the \emph{functional} execution from the \emph{timing} in most of its models.
For instance, each of gem5's CPU models can be used with any ISA as the ISA's functional implementation is separate from the CPU timing model.
This separation allows gem5 to create checkpoints during execution, fast-forward to the region of interest using low fidelity models, and introspection from the simulator into the simulated system at runtime.

\begin{figure*}
  \begin{subfigure}{0.28\linewidth}
    \centering
    \includegraphics[width=\linewidth]{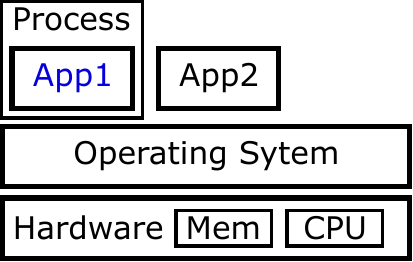}
    \caption{The common hardware/software abstraction layers.}
    \label{fig:gem5-fs-normal}
  \end{subfigure}
  \hfill
  \begin{subfigure}{0.28\linewidth}
    \centering
    \includegraphics[width=\linewidth]{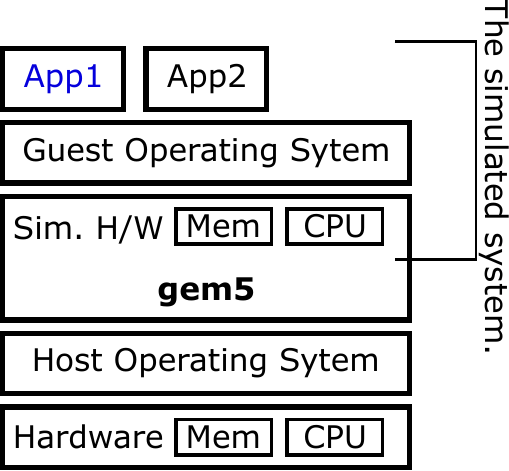}
    \caption{The hardware/software abstraction layers when using gem5 in full system simulation mode.}
    \label{fig:gem5-fs-fs}
  \end{subfigure}
  \hfill
  \begin{subfigure}{0.28\linewidth}
    \centering
    \includegraphics[width=\linewidth]{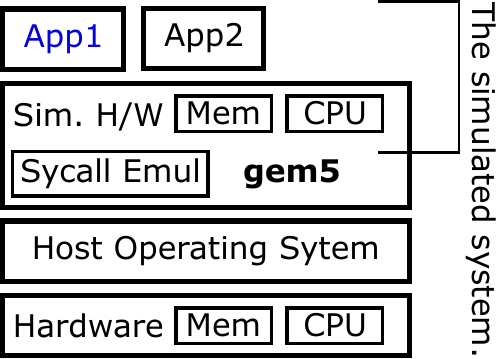}
    \caption{The hardware/software abstraction layers when using gem5 in system call emulation mode.}
    \label{fig:gem5-fs-se}
  \end{subfigure}
  \caption{A comparison of gem5's different modes of operation.}
  \label{fig:gem5-fs}
\end{figure*}

The gem5 simulator can be used in two different modes: full system simulation or system call emulation (syscall emul or SE-mode).
Figure~\ref{fig:gem5-fs} shows the hardware/software abstraction layers in each of these cases.
In \emph{full system} mode, gem5 can boot a full Linux-based operating system (e.g., Ubuntu 20.04).
After booting the guest OS, the researcher can run the application of interest to generate statistics.
In system call emulation mode, the gem5 simulator itself emulates the operating system.
The support for Linux system calls has been greatly improved recently (see Section~\ref{sec:se-mode}).
SE-mode ignores the timing of many system-level effects including system calls, TLB misses, and device accesses.
Thus, researchers should use caution when running experiments in SE-mode and ensure that ignoring system-level effects will not change the results of the experiments.


\subsubsection{gem5 Design}

\begin{figure*}
  \centering
  \includegraphics[width=\textwidth]{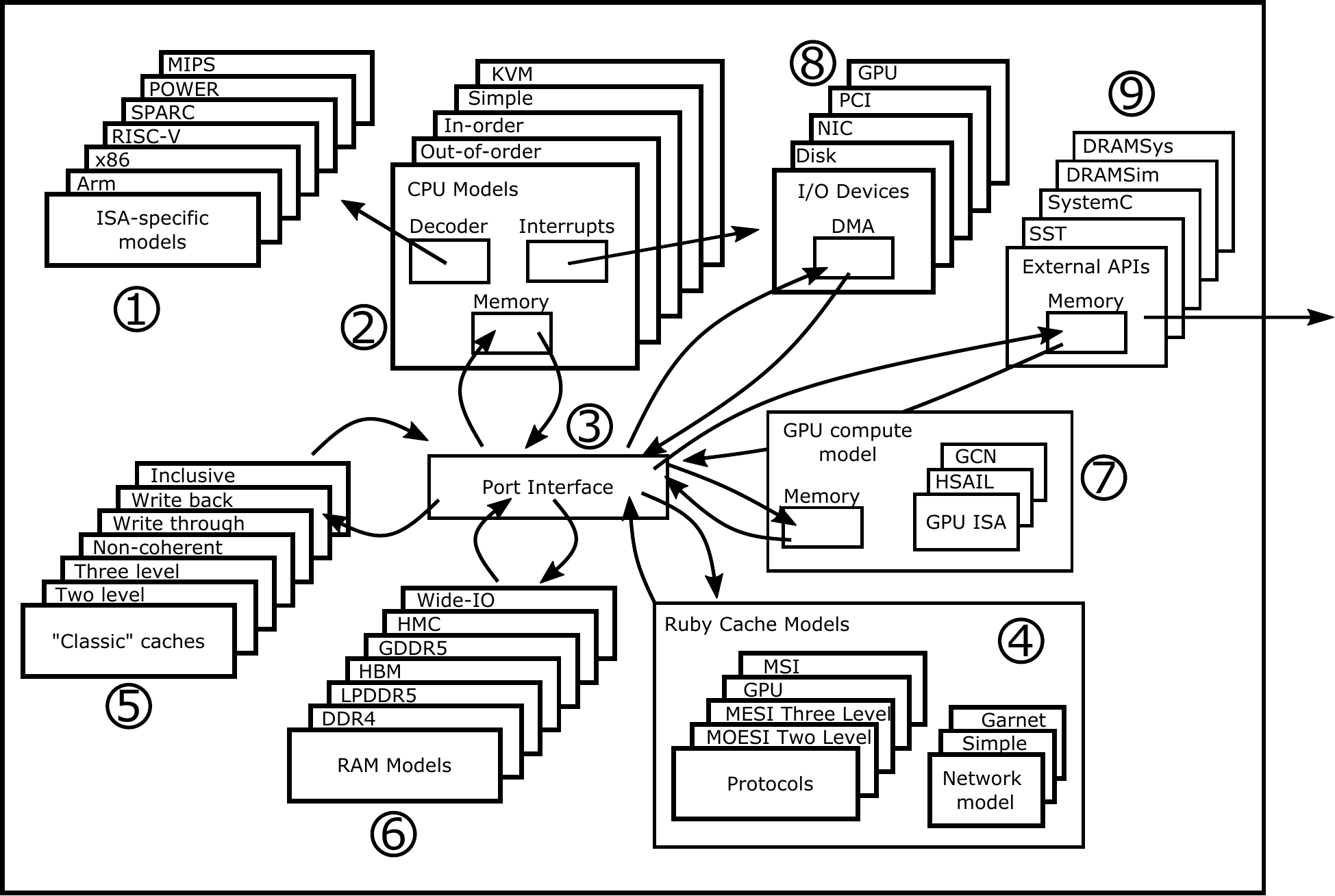}
  \caption{An overview of gem5's architecture. Its modular components allow any of each model type to be used in system configuration via Python scripts. Users can choose the fidelity of the memory system, CPU model, etc. while being able to select any ISA, devices, etc. The port interface allows any memory component to be connected to any other memory component as specified by the Python script. Details of each of these simulator components are discussed in Section~\ref{sec:main-features}}
  \label{fig:gem5-big-picture}
\end{figure*}

The gem5 simulator is a cycle-level computer system simulation environment.
At its core, gem5 contains an event-driven simulation engine.
On top of this simulation engine, gem5 implements a large number of models for system components including CPU cores (out-of-order designs, in-order designs, and others), a detailed DRAM model, on-chip interconnects, coherent caches, I/O devices, and many others.
All of these models are parameterized, and they can be customized for different systems (e.g., there are fully specified models for different DRAM devices including DDR3/4, GDDR5, HBM, and HMC, users can change the width of the out-of-order core, etc.).
Many of these different models are shown in Figure~\ref{fig:gem5-big-picture}.
The gem5 project also contains tests to help find bugs, a complex and feature-rich statistics database, and a Python scripting interface to describe systems under test and run simulations.

The gem5 simulator has modular support for multiple ISAs (see Figure~\ref{fig:gem5-big-picture}~\textcircled{1}).
The gem5 simulator currently supports Arm, GPU ISAs, MIPS, Power, RISC-V, SPARC, and x86.
These ISAs not only include the details to execute each instruction, but also the system-specific devices necessary for full system simulation.
There is robust full system support for Linux on Arm and x86.
Additionally, many other ISAs have some level of full system support.

All of these ISAs can be used with any of gem5's CPU models as the CPU models are designed to be ISA-agnostic (Figure~\ref{fig:gem5-big-picture}~\textcircled{2}).
Four different CPU models are included which span the fidelity-performance spectrum.
The gem5 simulator contains ``simple'' CPU models that can be used for memory system studies or other studies that do not require high fidelity execution models.
When using the ``simple'' CPU model, gem5 simulates the system faster, but has less fidelity when compared to real devices.
Additionally, gem5 contains a detailed in-order CPU model (the ``minor'' CPU) and an out-of-order CPU model (the ``O3'' CPU).
When using these high-fidelity models, gem5 runs slower, but can provide more realistic results.

Finally, gem5 includes a CPU ``model'' that bypasses simulation and allows the binaries running in gem5 to use the underlying host's processor, if the host ISA is the same as the application running in gem5.
In this mode, the performance of gem5 is nearly the same as when running natively~\cite{full-speed-ahead}.
This CPU model is based on the kernel virtual machine (KVM) API in Linux and leverages the hardware virtualization support available in many modern processors.
Although the KVM CPU model can execute at native speed, it does not model the timing of execution or memory requests.
The KVM-based CPU model can be used for sampled simulation and fast-forwarding to the region of interest and checkpoint locations.

To connect the different compute, memory, and I/O device models gem5 provides a modular port interface which allows any component that implements the port API to be connected to any other component implementing the same API (Figure~\ref{fig:gem5-big-picture}~\textcircled{3}).
This allows models designed for one system to be easily used in other system designs.

There are two different cache systems in gem5: Ruby (Figure~\ref{fig:gem5-big-picture} \textcircled{4}), which models cache coherence protocols with high fidelity; and the ``classic'' caches (Figure~\ref{fig:gem5-big-picture}~\textcircled{5}), which lack cache coherence fidelity and flexibility.
Ruby enables user-defined cache coherence protocols and gem5 includes many different protocols out of the box.
Users can also choose to use a simple network model or the detailed Garnet model~\cite{garnet-2} when using Ruby caches which offers cycle-level detail for the on-chip network.

The classic caches have a single hard-coded hierarchical MOESI coherence protocol.
However, this cache model is easily composable allowing users to construct hierarchical cache topologies without worrying about the details of the coherence protocol.
Both Ruby and the classic caches can be used with any CPU model, any ISA, and any memory controller model.

The gem5 simulator also includes an event-driven DRAM model (Figure~\ref{fig:gem5-big-picture}~\textcircled{6}).
This DRAM model is easily configurable with the timing parameters for a variety of different DRAM controllers including DDR3, DDR4, GDDR, HBM, HMC, LPDDR4, LPDDR5, and others.
Although this is not a cycle-accurate DRAM model like DRAMSim~\cite{wang_05, dramsim2, dramsim3} or Ramulator~\cite{yoongy_16}, it is nearly as accurate while providing more flexibility and higher performance~\cite{HanssonAgarwal2014-gem5DRAM}.

In addition to CPU models, gem5 also includes a cycle-level compute-based GPU~\cite{GutierrezBeckmann2018-amdAPU, Ta2019gputesting} (Figure~\ref{fig:gem5-big-picture}~\textcircled{7}).
This GPU model does not support graphics applications, but supports many compute applications based on the heterogeneous system architecture (HSA) and ROCm runtime.
The GPU model is based on AMD's Graphics Core Next (GCN) architecture~\cite{gcnWhitepaper, gcn3Manual}.
The GPU model has a modular ISA similar to the CPU model in gem5, and can be extended to support other GPU ISAs in the future. Additionally, gem5 contains support for a functional-only GPU model to enable simulating applications that depend on graphics APIs but do not depend on graphics performance~\cite{nomali}.

An important component to full system simulation is supporting I/O and other devices (Figure~\ref{fig:gem5-big-picture}~\textcircled{8}).
Thus, gem5 supports many system-agnostic devices such as disk controllers, PCI components, Ethernet controllers, and many more.
There are also many system-specific device models such as the Arm GIC and SMMU, and x86 PC devices.

Finally, gem5 has been integrated with other computer architecture simulator systems to enable users with models in other simulator systems to use gem5's features (Figure~\ref{fig:gem5-big-picture}~\textcircled{9}).
For instance, gem5 has been integrated with the Structural Simulation Toolkit (SST)~\cite{RodriguesHemmert2011-sst, HsiehPedretti2012-sst-gem5}, which uses gem5's detailed CPU models in conjunction with SST's multi-node modeling capabilities.
DRAMSim~\cite{wang_05, dramsim2, dramsim3} which provides cycle-accurate DRAM models has also been integrated with gem5.
Additionally, the IEEE standard SystemC API~\cite{menard2017-system-systemc} has been integrated to enable users with SystemC models to use them as gem5 components (see Section~\ref{sec:systemc} for more details).
One example for a coupling with a SystemC model is the flexible and cycle accurate DRAM subsystem design space exploration framework DRAMSys~\cite{junwei_13, junwei_15, stejun_20}, which is a based on SystemC TLM-2.0.

\section{Major Changes in gem5-20}
\label{sec:changes}

In addition to the systematic changes in project management discussed in Section~\ref{sec:current-gem5} there have also been many added features, fixed bugs, and general improvements to the codebase.
This section contains descriptions of some of the major changes to gem5.
There were 7015 commits by at least 250 unique contributors between when gem5 was released in 2011 and the release of gem5-20.
This section is a comprehensive, but not exhaustive, list of the major changes in gem5.
Along with the description of the changes in gem5, we also recognize the individuals or groups who made significant contributions to each of these features with separate by-lines for each subsection.
However, there are many unlisted contributors that were indispensable in getting gem5 where it is today.

Table~\ref{table:overview} gives and overview of the major changes in gem5 with pointers to subsections which contain more detail on each change.

\begin{table}
    \begin{tabular}{|p{0.95\linewidth}|}
        \hline
        \multicolumn{1}{|c|}{\textbf{General usability improvements} } \\
        \hline
        \textbf{Section~\ref{sec:resources}:} Added new resources repository with disk images, kernel images, etc. \\
        \textbf{Section~\ref{sec:learning}:} Learning gem5 book and class. Provides a way to get started using and developing with gem5. \\
        \hline \hline
        \multicolumn{1}{|c|}{ \textbf{ISA improvements} } \\
        \hline
        \textbf{Section~\ref{sec:riscv}:} RISC-V ISA added. User mode fully supported. Some support for full system. \\
        \textbf{Section~\ref{sec:arm}:} Arm ISA improvements. Added support for Armv8, SVE instructions, and trusted firmware. \\
        \textbf{Section~\ref{sec:x86}:} x86 ISA improvements. Better support for out-of-order CPU models, many instructions added, and support for TSO memory consistency. \\
        \hline \hline
        \multicolumn{1}{|c|}{\textbf{Execution model improvements} } \\
        \hline
        \textbf{Section~\ref{sec:predictor}:} New branch predictors including L-TAGE.\\
        \textbf{Section~\ref{sec:virtualized-ff}:} New CPU model based on KVM added. Uses host hardware to accelerate simulator. \\
        \textbf{Section~\ref{sec:elastic}:} Elastic trace execution added. Trace capture and playback with dynamic dependencies for fast flexible simulation. \\
        \hline \hline
        \multicolumn{1}{|c|}{\textbf{Memory system improvements} }\\
        \hline
        \textbf{Section~\ref{sec:dramcontroller}:} Configurable DRAM controller. Added support for many DRAM devices, low-power DRAM, quality of service, and power models. \\
        \textbf{Section~\ref{sec:classic}:} Classic cache improvements. Added non-coherent caches, write streaming optimizations, cache maintenance operations, and snoop filtering. \\
        \textbf{Section~\ref{sec:replacement}:} General replacement policy framework and cache compressions support added. \\
        \textbf{Section~\ref{sec:ruby}:} Ruby model improvements. Many general improvements, GPU coherence protocols, and support for Arm ISA. \\
        \textbf{Section~\ref{sec:garnet}:} Garnet network improved to version 2.0 with more detailed router and network models. \\
        \hline \hline
        \multicolumn{1}{|c|}{\textbf{New models} } \\
        \hline
        \textbf{Section~\ref{sec:gpu}:} GPU compute model added. Models AMD's GCN architecture in SE-mode with support for shared memory systems. Tests for GPU-like coherence protocols also added. \\
        \textbf{Section~\ref{sec:dvfs}:} Runtime power modeling and DVFS support added. \\
        \textbf{Section~\ref{sec:virtio-nomali}:} Support for timing-agnostic device models added. VirtIO enables more flexible guest-simulator interaction and the NoMali GPU model allows graphic-based applications to execute more realistically. \\
        \textbf{Section~\ref{sec:dist-gem5}:} Support for modeling multiple distributed systems added. \\
        \textbf{Section~\ref{sec:systemc}:} SystemC model integration. Added a bridge from SystemC TLM models to gem5 models, and added an implementation of SystemC for tight gem5-SystemC integration. \\
        \hline \hline
        \multicolumn{1}{|c|}{\textbf{General infrastructure improvements} } \\
        \hline
        \textbf{Section~\ref{sec:se-mode}:} SE-mode improvements. Support for dynamically-linked binaries, more system calls, multi-threaded applications, and a virtual file system. \\
        \textbf{Section~\ref{sec:testing}:} Testing improvements. New unit test framework and continuous integration support. \\
        \textbf{Section~\ref{sec:internal}:} General infrastructure improvements. Added support for HD5F output for statistics, Python 3 support, and asynchronous modeling. \\
        \textbf{Section~\ref{sec:guest-sim}:} Updated guest$\leftrightarrow$simulator APIs. \\
        \hline
    \end{tabular}
    \caption{Overview of major change in gem5.}
    \label{table:overview}
\end{table}

\subsection[gem5 resources]{gem5 Resources\footnote{by Ayaz Akram, Bobby R. Bruce, Hoa Nguyen, and Mahyar Samani}}
\label{sec:resources}

The gem5 simulator permits the simulation of many different systems
using a variety of benchmarks and tests.
However, gathering and compiling the resources to perform experiments with gem5 can be a laborious process.
To provide a better user-experience we have began maintaining \emph{gem5 resources}, which we broadly define as a set of artifacts that are not required to build or run gem5, but that may be utilized to carry out experiments and simulations.
For example, Linux kernels, disk images, popular benchmark suites, and commonly used tests binaries are frequently needed by users of gem5 but are not distributed as part of the gem5 itself.
As part of our gem5-20 release, these resources, with source code and build instructions for each, are gradually being centralized in a common repository\footnote{\url{https://gem5.googlesource.com/public/gem5-resources}}.

A key goal of this repository is to ensure reproducibility of gem5 experiments.
The gem5 resources repository provides researchers with a suite of disk images with pre-installed operating systems and benchmarks as well as kernel binaries.
Thus, all researchers which use the resources are starting from a common point and can more easily reproduce experiments.
Additionally, all of the sources and scripts to build each artifact are also included in the repository which can be modified to create custom resources.

\subsubsection{Testing gem5-20 with gem5 Resources}

Another important aim of creating a common set of gem5 resources is to more regularly test gem5 on a suite of common benchmarks, operating systems Linux kernels.
As part of gem5-20, we have tested the simulator's capability to run SPEC 2006~\cite{spec06}, SPEC 2017~\cite{spec17}, PARSEC~\cite{parsec}, the NAS Parallel Benchmarks (NPB)~\cite{npb}, and the GAP Benchmark Suite (GAPBS)~\cite{gapbs}.
We have also shown gem5-20's performance when running five different long-term service (LTS) Linux kernel releases on a set of different CPU and memory configurations.
The results from these investigations can be found on our website~\footnote{\url{http://www.gem5.org/documentation/benchmark_status}}.
We plan to use this information, and gem5 resources repository, to better target problem areas in the gem5 project.

Furthermore, with a shared set of common resources and knowledge of what configurations work best with gem5, we can provide the community with a set of ``known good'' gem5 configurations to facilitate computer architecture research.
We intend for these configurations to replicate the functionality and performance of architectural components at a high level of fidelity.

\subsection[Learning gem5]{Learning gem5\footnote{By Jason Lowe-Power}}
\label{sec:learning}

The gem5 simulator has a steep learning curve.
Often, using gem5 in research means \emph{developing} the simulator to modify or add new models.
Not only do new users have to navigate the hundreds of different models, but they also have to understand the core of the simulation framework.
We found that this steep learning curve was one of the biggest impediments to productively using gem5.
There was anecdotal evidence that it would take new users \emph{years} to learn to use gem5 effectively~\cite{Power-gem5horrors-2015}.
Additionally, due to a lack of formal documentation, the only way to learn parts of gem5 was to work with a senior graduate student or to intern at a company and pick up the knowledge ``on the job''.

\emph{Learning gem5} reduces the knowledge gap between new users and experienced gem5 developers.
\emph{Learning gem5} takes a bottom-up approach to teaching new users the internals of gem5.
There are currently three parts of \emph{Learning gem5}, ``Getting Started'', ``Modifying and Extending'', and ``Modeling Cache Coherence with Ruby''.
Each part walks the reader through a step-by-step coding example starting from the simplest possible design up to a more realistic example.
By explaining the thought process behind each step, the reader gets a similar experience to working alongside an experienced gem5 developer.
\emph{Learning gem5} includes documentation on the gem5 website\footnote{\url{http://www.gem5.org/documentation/learning_gem5/introduction/}} and source code in the gem5 repository for these simple ground-up models.

Looking forward, we will be significantly expanding the areas of the simulator covered by \emph{Learning gem5} and creating a gem5 ``summer school'' initially offered summer of 2020.
This ``summer school'' will mainly be an online class (e.g., Coursera) with all videos available on the gem5 YouTube channel\footnote{\url{https://www.youtube.com/channel/UCCpCGEj_835WYmbB0g96lZw}}, but we hope to have in-person versions of the class as well.
These classes will also be the basis of gem5 Tutorials held with major computer architecture and other related conferences.

\subsection[RISC-V ISA Support]{RISC-V ISA Support}
\label{sec:riscv}

RISC-V is a new ISA which has quickly gained popularity since its creation in 2010, only one year before the initial gem5 release~\cite{Waterman2011riscv}.
In this time, the number of RISC-V users has grown significantly, especially in the computer architecture research community.
Thus, the addition of RISC-V as a supported ISA for gem5 is one of the main new features in the past nine years.

\subsubsection[General RISC-V ISA Implementation]{General RISC-V ISA Implementation\footnote{By Alec Roelke}~\cite{risc5-gem5, risc5-multicore-gem5}}

The motivation for implementing the RISC-V ISA into gem5 stemmed from needing a way to explore architectural parameters for RISC-V designs.
At the time of implementation, the only means of simulating RISC-V was using Spike (its simplified, single-cycle RTL simulator), QEMU, full RTL simulation, or emulation on an FPGA.
Spike and QEMU are not detailed enough and RTL simulation is too time consuming for these methods to be feasible for architectural parameter exploration.
With FPGA emulation, it is difficult to retrieve performance information without modifying both the RTL design and the system software.
The gem5 simulator provides an easy means of performing architectural analysis through its detailed hardware models.

The implementation follows the divisions of the instruction set into its base ISA and extensions, beginning with the 32-bit integer base set, RV32I.
This implementation was modeled off of the existing gem5 code for MIPS and Alpha ISAs, which are also RISC instruction sets that share many of the same operations as RISC-V.
Including support for 64-bit addresses and data (RV64) and for the multiply (M) extension mainly involved adding the new instructions and changing some parameters to expand register and data path widths.

The next two extensions, atomic (A) and floating point (F and D for single- and double-precision, respectively), were more complicated.
The A extension includes both load-reserved/store-conditional (LR/SC) sequence of instructions for performing complex atomic operations on memory and a set of read-modify-write instructions for performing simple ones.
These instructions were implemented as a pair of micro-ops that acted like an LR/SC pair with one of the pair additionally performing the specified operation.
Floating-point instructions required many special cases to ensure correct error handling and reporting, and we were not able to implement one of the five possible rounding modes (round away from zero) RISC-V specifies for inexact calculations due to the fact that C++ does not support it.
Finally, support for the non-standard compressed (C) extension, which adds 16-bit versions of high-usage instructions, was added when it was discovered that this extension was included by default in many RISC-V software toolchains (e.g., GCC).
The compressed instruction implementation required creating a state machine in the instruction decoder to keep track of whether the current instruction is compressed, to increment the PC by the correct amount based on the size of the instruction, and to handle cases where a full-length instruction crosses a 32-bit word boundary.

With this implementation, most RISC-V Linux programs are supported in system call emulation mode.
Continued work has improved the implementation of atomic instructions, including actual atomic read-modify-write accesses in a single instruction and steps toward support for full system simulation.
Additionally, gem5's version of the RISC-V test-suite\footnote{\url{https://github.com/riscv/riscv-tests}} has been updated to the latest version and several corner cases in gem5 have been fixed, so that now most of the tests are working correctly.

\subsubsection[RISC-V Full System Support]{RISC-V Full System Support\footnote{By Nils Asmussen}}

To simulate complete operating systems the RISC-V ISA has been extended to support full system simulation.
More specifically, we added support for Sv39 paging according to the privileged ISA 1.11\footnote{\url{https://riscv.org/specifications/privileged-isa/}} with a 39-bit virtual address space, a page-table walker performing a three-level translation, and a translation lookaside buffer (TLB).
The page-table walker code is based on the existing gem5 code for x86 due to the structural similarities.
While a few steps are still missing to run Linux, general support to run a complete RISC-V operating system on gem5 is available now.

\subsection[Arm Improvements]{Arm Improvements}
\label{sec:arm}

\subsubsection[Armv8-A Support]{Armv8-A Support\footnote{by Giacomo Gabrielli, Andreas Sandberg, and Giacomo Travaglini}}

The Armv8-A architecture introduced two different architectural states:
AArch32, supporting the A32 and T32 instruction sets (backward-compatible with
Armv7-A and Thumb instruction sets, respectively), and AArch64, a new
state offering support for 64-bit addressing via the A64 instruction set. Currently, gem5
supports all of the above instruction sets and the interworking
between them.
On top of the user-level features, several important system-level extensions, e.g. the
security (aka TrustZone\textregistered~\cite{ArmTustZone}) and virtualization extensions~\cite{ArmARM}, have been contributed opening up new avenues for architectural and microarchitectural research.

While Armv8-A was a major iteration of the architecture, there have been
several smaller iterations introduced by Arm with a yearly cadence, and various
contributors have implemented some of the main features from those extensions,
up to Armv8.3-A.

\subsubsection[Support for the Arm Scalable Vector Extension (SVE)]{Support for the Arm Scalable Vector Extension (SVE)\footnote{by Gabor Dozsa, Giacomo Gabrielli, Rekai Gonzalez-Alberquilla, Nathanael Premillieu, and Javier Setoain}}

In 2016, Arm introduced their Scalable Vector Extension (SVE)~\cite{ArmARM}, a
novel approach to vector instruction sets. Instead of having fixed-size vector
registers, SVE operates on registers that can be anywhere between 128 to 2048
bit long (in 128-bit increments). SVE code is arranged in a way that is agnostic to the
underlying vector length (Vector Length Agnostic Programming), and a single SVE
instruction will perform its operation on as many elements as the vector
register can fit, depending on its length. On top of the 32 variable-length
vector registers, SVE also adds 16 variable length predicate registers for
predicated execution. These registers store one bit per byte (the minimum
element size) in the vector register, and can be used to select specific
elements in the vector for operation~\cite{white-paper-on-SVE-and-VLA-programming}.

To support SVE, gem5 implements register storage and register access
as two separated classes, a container and an interface, decoupling one from the
other. The vector registers can be of any arbitrary size and be accessed as
vectors of elements of any particular type, depending on the operand types of
each instruction. This not only facilitates handling variable size registers,
it also abstracts the nuances of handling predicate registers, where the stored
values have to be grouped and interpreted differently depending on the operand
type.

This design provides enough flexibility to support any vector instruction sets
with arbitrarily large vector registers.

\subsubsection[Trusted Firmware Support]{Trusted Firmware Support\footnote{by Adrian Herrera}}

Trusted Firmware (TF-A) is Arm's reference implementation of Secure World software for A-profile architectures.
It enables Secure Boot flow models, and provides implementations for the Secure Monitor executing at Exception Level 3 (EL3) as well as for several Arm low-level software interface standards, including System Control and Management Interface (SCMI) driver for accessing System Control Processors (SCP), Power State Coordination Interface (PSCI) library support for power management, and Secure Monitor Call (SMC) handling.

TF-A is supported on multiple Arm Development Platforms (APDs), each of them characterized by its set of hardware components and their location in the memory map (e.g., Juno ADP and the Fixed Virtual Platforms (FVP) ADP family).
However, the Arm reference platforms in gem5 are part of the \lstinline|VExpress_GEM5_Base| family.
These are loosely based on a Versatile\texttrademark Express RS1 platform with a slightly modified memory map. TF-A implementations are provided for both Juno and FVPs, however not for \lstinline[breaklines,breakatwhitespace]|VExpress_GEM5_ Base|.

Towards unifying Arm's platform landscape, we now provide a \lstinline|VExpress_GEM5_Foundation| platform as part of gem5's \lstinline[breaklines,breakatwhitespace]|VExpress_ GEM5_Base| family.
This is based on and compatible with FVP Foundation, meaning all Foundation software may run unmodified in gem5, including but not limited to TF-A.
This allows for simulating boot flows based on UEFI implementations (U-boot, EDK II), and brings us a step closer to support for all UEFI compatible operating systems in gem5.

\subsection[X86 ISA Improvements]{X86 ISA Improvements\footnote{by Nilay Vaish}}
\label{sec:x86}

The x86 or x86-64 ISA is one of the most popular ISAs for desktop, server, and high-performance compute systems.
Thus, there has been significant effort to improve gem5's modeling of this ISA.
This section presents a subset of the changes to improve the x86 ISA.
There are many other improvements large and small that generally have improved the fidelity of x86 modeling.

In out-of-order CPUs (e.g., gem5's O3CPU), instructions whose dependencies have been satisfied are allowed to execute even if there are instructions earlier in the stream waiting for their operands.
The flag register used in the x86 ISA complicates this out-of-order execution as almost every instruction both reads and writes this register making them all dependent on one another.
Maintaining a single flag register can introduce dependencies that need not exist.
We now maintain multiple flag registers for holding subsets of flag bits to reduce the dependencies.
This prevents unnecessary serialization, unlocking a significant amount of instruction-level parallelism.

Memory consistency models decide the amount of parallelism available in a memory system, while correctly executing a program.
The x86 architecture is based on the Total Store Order (TSO) memory model~\cite{NagarajanSorin2020-cohMCMPrimer}.
We added support for TSO to gem5 for the x86 architecture.
This meant ensuring that a later load from a thread can bypass earlier loads and stores, but stores from the same thread are always executed in order.
The out-of-order CPU model in gem5 has been improved to implement both TSO and more relaxed consistency models (e.g., those in the RISC-V and Arm architectures discussed in Sections~\ref{sec:riscv} and \ref{sec:arm}, respectively).

\subsection[Branch Predictor Improvements]{Branch Predictor Improvements\footnote{by Dibakar Gope}}
\label{sec:predictor}

In gem5, multiple branch prediction models are available, many of which were added since the initial release of gem5.
Currently, gem5 supports five different branch prediction techniques including the well-known TAGE predictor as well as standard predictors such as bi-mode, tournament, etc.
This list can easily be expanded to cover different variants of these well-known branch predictors.
Besides, the support for loop predictor and indirect branch predictor is also available.

Furthermore, the modularity of the implementation of different branch predictors allows ease of inclusion of secondary or side predictors into the prediction mechanism of primary predictors.
For example, TAGE can be seamlessly augmented with a loop predictor to predict loops with constant iteration numbers.
Indirect branch predictor can be made to use complex TAGE-like scheme instead of simple history-based predictors with only a few hours of development effort.
In addition to this, these different predictors can be configured with different sizes of history registers and table-like structures.
For example, TAGE predictor can be configured to run with different sizes of the history register and consequently a different number of predictor tables, allowing users to investigate the effects of different predictor sizes in various performance metrics.

Future development is planned to include the support of neural branch predictors (e.g., perceptron branch predictor, etc.) and different variants of TAGE and perceptron predictors that have demonstrated significant improvement in branch misses in recent years.

\subsection[Virtualized Fast Forwarding]{Virtualized Fast Forwarding\footnote{by Andreas Sandberg}}
\label{sec:virtualized-ff}

Support for hardware virtualization (e.g., AMD-V, Intel VT-x, and Arm virtualization extensions) is a very useful feature for bring up, model development, testing, and novel simulation research~\cite{full-speed-ahead, NikolerisSHC16, NikolerisEHC19}.
Work on the original implementation of hardware virtualization support started in the summer 2012 in Arm Research and targeted the Arm Cortex A15 chip. Some of the most challenging parts of the development were the lack of a stable kernel API for KVM on Arm and the limited availability of production silicon.
However, despite these challenges, we had a working prototype that booted Linux in autumn.
This prototype was refined and merged into gem5 in April 2013, just one month after QEMU gained support for Arm KVM.
Support for x86 followed later that year.
A good overview of the KVM implementation can be found in the technical report by Sandberg et. al~\cite{full-speed-ahead}.
The original full-system implementation was later extended to support syscall emulation mode on x86~\cite{DutuSlice2015-kvm}.

Support for hardware virtualization in gem5 enabled research into novel ways of accelerating simulation.
The original intention was to use KVM to generate checkpoints and later simulate those checkpoints in parallel with different system parameters (e.g., to do design space exploration).
However, we quickly realized that the checkpointing step could be eliminated by cloning the simulator state at runtime.
This led to the introduction of the fork call in gem5's Python API.
Under the hood, this call drains the simulator to make sure everything is in a consistent state, it then uses the UNIX fork call to create a copy of the simulator.
A typical use case uses a main process that generates samples that are simulated in parallel.
More advanced use cases use fork semantics to simulate multiple outcomes of a sample to quantify the cache warming errors introduced by using KVM to fast-forward between samples~\cite{full-speed-ahead}.

\subsection[Elastic Traces]{Elastic Traces\footnote{by Radhika Jagtap, Matthias Jung, Stephan Diestelhorst, Andreas Hansson, Thomas Grass, and Norbert Wehn}}
\label{sec:elastic}

Detailed execution-driven CPU models, like gem5's out-of-order model, offer high accuracy, but at the cost of simulation speed.
Therefore, trace-driven simulations are widely adopted to alleviate this problem, especially for studies focusing on memory-system exploration.
However, traces with fixed time stamps always include the implicit behavior of the simulated memory system with which they were recorded.
If the memory system is changed during exploration this can lead to wrong simulation results, since an out-of-order core would react differently on the new memory system.
Ideally, trace-driven core models will mimic out-of-order processors executing full-system workloads by respecting true dependencies and ignoring false dependencies to enable computer architects to evaluate modern systems.

We implemented the concept of elastic traces in which we accurately capture data and memory order dependencies by instrumenting a detailed out-of-order processor model~\cite{jagdie_16}.
In contrast to existing work, we do not rely on offline analysis of timestamps, and instead use accurate dependency information tracked inside the processor pipeline.
We thereby account for the effects of speculation and branch misprediction resulting in a more accurate trace playback compared to fixed time traces.
We integrated a trace player in gem5 that honors the dependencies and thus adapts its execution time to memory-system changes, as would the actual CPU.
Compared to the detailed out-of-order CPU model, our trace player achieves a speed-up of 6-8 times while maintaining a high simulation accuracy (83--93\%), achieving fast and accurate system performance exploration.

\subsection[Off-Chip Memory System Models]{Off-Chip Memory System Models\footnote{by Nikos Nikoleris}}
\label{sec:dramcontroller}

The gem5 simulator can model a large number of configurations in the off-chip memory system.
Its memory controller handles requests from the on-chip memory system and issues read and write commands to the actual memory device, modeling the timing behavior of the latter.
Over the years a number of contributions have added features that allow modeling emerging new technologies and features as documented below.

\subsubsection[New memory controller features]{New memory controller features\footnote{by Wendy Elsasser}}

The gem5 DRAM controller provides the interface to external memory, which is traditionally DRAM.
It consists of two main components: the memory controller itself and the DRAM interface.
The DRAM interface contains media specific information, defining the architecture and timing parameters of the DRAM as well as the functions that manage the media specific operations like activation, precharge, refresh and low power modes~\cite{HanssonAgarwal2014-gem5DRAM}.
These models are easily modified by extending a Python class and updating the timing parameters for a new DRAM device.

\subsubsection[Low-power DDR]{Low-power DDR\footnote{by Wendy Elsasser}}

LPDDR5 is currently in mass production for use in multiple markets including mobile, automotive, AI, and 5G.
This technology is expected to become the mainstream Flagship Low-Power DRAM by 2021 with anticipated longevity due to proposed speed grade extensions.
The specification defines a flexible architecture and multiple options to optimize across different use cases, trading off power, performance, reliability and complexity.
To evaluate these tradeoffs, we have updated the memory controller to support the new features and added LPDDR5 configurations.

While these changes have been incorporated for LPDDR5, some of them could be applicable to other memory technologies as well.
The gem5 changes incorporate new timing parameters, support for multi-cycle commands, and support for interleaved bursts.
These features require new checks and optimizations in gem5 to ensure the model integrity when comparing to real hardware.
For example, support for multi-cycle commands along with the changes to LPDDR5 clocking motivated a new check in gem5 to verify command bandwidth.
Previously, the DRAM controller did not verify contention on the command bus and assumed adequate command bandwidth, but with the evolution of new technologies this assumption is not always valid.

\subsubsection[Quality of Service Extensions]{Quality of Service Extensions\footnote{by Matteo Andreozzi}}

The coexistence of heterogeneous tasks/workloads on a single computer system is common practice in modern systems, from the automotive to the high-performance computing use-case.
Quality of Service (QoS) is the ability of a system to provide differential treatment to its clients, in a quantifiable and predictable way.



We now include a QoS-aware memory controller in gem5, and the definition of basic (example) policies modeling the prioritization algorithm of the memory controller.
We include models for a \emph{fixed} priority policy (every requestor in the system has a fixed priority assigned) and the \emph{proportional fair} policy (where the priority of a requestor is dynamically adjusted at runtime based on utilization).

The default timing-based DRAM controller described above has been rewritten to include the QoS changes.
These changes separate out the QoS policy from the DRAM timing model.
With the framework in place a user can write its own policy and seamlessly plug it into a real memory controller model to unlock system wide explorations under its own arbitration algorithm.

\subsubsection[DRAMPower and DRAM Power-Down Modes]{DRAMPower and DRAM Power-Down Modes\footnote{by Matthias Jung, Wendy Elsasser, Radhika Jagtap, Subash Kannoth, Omar Naji, \'Eder F.
Zulian, Andreas Hansson, Christian Weis, and Norbert Wehn }}
Across applications, DRAM is a significant contributor to the overall system power.
For example, the DRAM access energy per bit is up to three orders of magnitude higher compared to an on-chip memory access.
Therefore, an accurate and fast power estimation is crucial for an efficient design space exploration.
DRAMPower~\cite{kargoo_14} is an open source tool for fast and accurate power and energy estimation for several DRAM memories based on JEDEC standards.
It supports unique features like power-down, bank-wise power estimation, per bank refresh, partial array self-refresh, and many more.
In contrast to Micron's DRAM Power estimation spread sheet\footnote{\url{https://www.micron.com/support/tools-and-utilities/power-calc}}, which estimates the power from device manufacturer's data sheet and workload specifications (e.g. Rowbuffer-Hit-Rate or Read-Write-Ratio), DRAMPower uses the actual timings from the memory transactions, which leads to a much higher accuracy in power estimation.
Furthermore, the DRAMPower tool performs DRAM command trace analysis based on memory state transitions and hence avoids cycle-by-cycle evaluation, thus speeding up simulations.

For the efficient integration of DRAMPower into gem5, we changed the tool from a standalone simulator to a library that could be used in discrete event-based simulators for calculating the power consumption online during the simulation.
Furthermore, we integrate the power-down modes into the DRAM controller model of gem5~\cite{jagjun_17} in order to provide the research community a tool for power-down analysis for a breadth of use cases. We further evaluated the model with real HPC workloads, illustrating the value of integrating low power functionality into a full system simulator.
\subsubsection[Future Improvements to Off Chip Memory Models]{Future Improvements to Off Chip Memory Models\footnote{by Wendy Elsasser}}
\label{sec:nvm}

We are currently working to refactor the DRAM interface to be extensible and enable modeling of other memory devices.
For instance, with the advent of SCM (storage class memory), emerging NVM (Non-Volatile Memory) could also exist on a memory interface, potentially alongside DRAM.
To enable support of NVM and future memory interfaces, we use a systematic approach to refactor the DRAM controller.
We pull the DRAM interface out of the controller and moved to a separate DRAM interface object.
In parallel, we create an NVM interface to model an agnostic interface to emerging memory.

The DRAM interface and the NVM interface have configurable address ranges allowing flexible heterogeneous memory configurations.
For example, a single memory controller can have a DRAM interface, an NVM interface, or both interfaces defined.
Other configurations are feasible, providing a flexible framework to study new memory topologies and evaluate the placement of emerging NVM in the memory sub-system.

\subsection[Classic Caches Improvements]{Classic Caches Improvements\footnote{by Nikos Nikoleris}}
\label{sec:classic}

The classic memory system implements a snooping MOESI-like coherence protocol that allows for flexible, configurable cache hierarchies.
The coherence protocol is primarily implemented in the \lstinline|Cache| and the \lstinline|CoherentXBar| classes and the \lstinline|SnoopFilter| object implements a common optimization to reduce unnecessary coherence traffic.

Over the years, the components of the classic memory system have received significant contributions with a primary focus of adding support for future technologies and enhancing its accuracy.

\subsubsection[Non-Coherent Cache]{Non-Coherent Cache}
The cache model in gem5 implements the full coherence protocol and as a result can be used in any level of the coherent memory subsystem (e.g., as an L1 data cache or instruction cache, last-level cache, etc.).
The non-coherent cache is a stripped down version of the cache model designed to be used below the point-of-coherence (closer to memory).
Below the point-of-coherence, the non-coherent cache receives only requests for fetches and writebacks and itself send requests for fetches and writebacks to memory below.
As opposed to the regular cache, the non-coherent cache will not send any snoops to invalidate or fetch data from caches above.
As such the non-coherent cache is a greatly simplified version in terms of handling the coherence protocol compared to the regular cache while otherwise supporting the same flexibility (e.g., configurable tags, replacement policies, inclusive or exclusive, etc.).

The non-coherent cache can be used to model system-level caches, which are often larger in size and can be used by CPUs and other devices in the system.

\subsubsection[Write Streaming Optimizations]{Write Streaming Optimizations}

Write streaming is a common access pattern which is typically encountered when software initializes or copies large memory buffers (e.g., memset, memcpy).
When executed, the core issues a large number of write requests to the data cache. The data cache receives these write requests and issues requests for exclusive copies of the corresponding cache lines. To get an exclusive copy, it has to invalidate copies of that line and fetch a copy of the data (e.g., from off-chip memory). As soon as it receives data, it performs all writes for that line and often will overwrite it completely. As a result, the data cache unnecessarily fetches data only to overwrite it shortly after. Often these write buffers are large in size and also trash the data cache.

Common optimizations~\cite{10.1145/173682.165154} coalesce writes to form full cache line writes, avoid unnecessary data fetches and achieve significant reduction in memory bandwidth.
In addition, when the written memory buffer is large, we can also avoid thrashing the data cache by bypassing allocation.

We have implemented a simple mechanism to detect write streaming access patterns and enable coalescing and bypassing.
The mechanism attaches to the data cache and analyses incoming write requests. When the number of sequential writes reaches a first threshold, it enables write coalescing and when a second threshold is reached, in addition, the cache will bypass allocation for the writes in the stream.

\subsubsection[Cache Maintenance Operations]{Cache Maintenance Operations}

Typically, the contents of the cache are handled by the coherence protocol.
For most user-level code, caches are invisible.
This greatly simplifies programming and ensures software portability.
However, when interfacing with devices or persistent memory, the effect of caching becomes visible to the programmer.
In such cases, a user might have to trigger a writeback which propagates all the way to the device or the persistent memory.
In other cases, a cache invalidation will ensure that a subsequent load will fetch the newest version of the data from a buffer of the main memory.

Cache maintenance operations (CMOs) are now supported in gem5 in a way that can deal with arbitrary cache hierarchies.
An operation can either clean and/or invalidate a cache line.
A clean
operation will find the dirty copy and trigger a writeback and an invalidate operation will find all copies of the cache line and invalidate them and the combined operation will perform both actions.
The effects of CMOs are defined with reference to a configurable point in the system.
For example, a clean and invalidate sent to the point-of-coherence will find all copies of the block above the point-of-coherence, invalidate them, and if any of them is dirty also trigger a writeback to the memory below the point-of-coherence.

\subsubsection[Snooping Support and Snoop Filtering]{Snooping Support and Snoop Filtering}

In large systems, broadcasting snoop messages is slow, they cost energy and time, and they can cause significant scalability bottlenecks.
Therefore, snoop filters (also called directories) are used to keep track of which caches or nodes are keeping a copy of a particular cached line.
We added a snoop filter to gem5 which is a distributed component that keeps track of the coherence state of all lines cached ``above'' it, similar to the AMD Probe Filter~\cite{Conway:opteron:2010}.
For example, if the snoop filter sits next to the L3 cache and is accessed before the L3, it knows about all lines in the L2 and L1 caches that are connected to that L3 cache.

Using the snoop filter, we can reduce the amount of messages from $O(N^2)$ to $O(N)$ with $N$ concurrent requestors in the system.
Modeling the snoop filter separately from the cache allows us to use different organizations for the filter and the cache, and distributing area between shared caches vs coherence tracking filters.
We also model the effect of limited filter capacity through back-invalidations that remove cache entries if the filter becomes full for more realistic cache performance metrics.]s
Finally, the more centralized coherence tracking in the filter allows for better checking of correct functionality of the distributed coherence protocol in the classic memory system.

\subsection[Cache Replacement Policies and New Compression Support]{Cache Replacement Policies and New Compression Support\footnote{By Daniel Rodrigues Carvalho}}
\label{sec:replacement}

In general, hardware components frequently contain tables whose contents are managed by replacement policies.
In gem5, multiple replacement policies are available, which can be paired with any table-like structure, allowing users to carry out research on the effects of different replacement algorithms in various hardware units.
Currently, gem5 supports 13 different replacement policies including several standard policies such as LRU, FIFO, and Pseudo-LRU, and various RRIPs~\cite{Jaleel2010rrip}.
These policies can be used with both the classic caches and Ruby caches.
This list is easily expandable to cover schemes with greater complexity as well.

These replacement policies are a great example of gem5's modularity and how code developed for one purpose can be reused in many other parts of the simulator.
Current and future development is planned to increase the use of these flexible replacement policies.
For instance, we are planning to extend the TLB and other cache structures beyond the data caches to take advantage of the same replacement policies.

The simulator also supports cache compression by providing several state-of-the-art compression algorithms~\cite{sardashti2015primer} and a default compression-oriented cache organization.
This basic organization scheme is derived from accepted approaches in the literature: adjacent blocks share a tag entry, yet they can only be co-allocated in a data entry if each block compresses to at least a specific percentage of the cache line size.
Currently, only BDI~\cite{pekhimenko2012base}, C-Pack~\cite{chen2010c}, and FPCD~\cite{alameldeen2018opportunistic} are implemented, but the modularity of the compressors allows for simple implementation of other dictionary-based and pattern-based compression algorithms.
Although these cache compression policies have only been applied to the classic caches, we are planning to use the same modular code to enable cache compression for the Ruby caches as well.

\subsection[Ruby Cache Model Improvements]{Ruby Cache Model Improvements}
\label{sec:ruby}

The Ruby cache model, originally from the GEMS simulator~\cite{MartinSBMXAMHW05}, is one of the key differentiating features of gem5.
The domain-specific language SLICC allows users to define new coherence protocols with high fidelity.
In mainline gem5, there are now 12 unique protocols including GPU-specific protocols, region-coherence protocols~\cite{PowerBasu2013-hsc}, research protocols like token coherence~\cite{MartinHill2003-tokenCoh}, and teaching protocols~\cite{NagarajanSorin2020-cohMCMPrimer}.

When gem5 was first released, Ruby had just been integrated into the project.
In the nine years since, Ruby and the SLICC protocols have become much more deeply integrated into the general gem5 memory system.
Today, Ruby shares the same replacement protocols (Section~\ref{sec:replacement}), the same port system to send requests into and out of the cache system, and the same flexible DRAM controller models (Section~\ref{sec:dramcontroller}) as the classic cache models.

Looking forward, we will be further unifying the Ruby and classic cache models.
Our goal is to one day have a unified cache model which has the composability and speed of the classic caches and the flexibility and fidelity of SLICC protocols.

\subsubsection[General Improvements]{General Improvements\footnote{by Nilay Vaish}}

Ruby now supports state checkpointing and restoration with warm cache.
This enables running simulations from regions of interest, rather than having to start fresh every time.
To enable checkpoints, we support accessing the memory system functionally i.e. without any notion of time or events.
The absence of timed events allows much higher simulation speeds.

Additionally, a new three level coherence protocol (\lstinline[breaklines,breakatwhitespace]|MESI_Three_ Level|) has been added to gem5.
For simplicity, this protocol was built on top of a prior two level protocol by adding an ``zero level'' (L0) cache at the CPU cores.
At the L0, the protocol has separate caches for instructions and data.
The L1 and the L2 caches are unified and do not distinguish between instructions and data.
The L0 and L1 caches are private to each CPU core while the L2 is shared across either all cores or a subset.



\subsubsection[Arm Support and Extensions]{Arm Support in Ruby Coherence Protocols\footnote{by Tiago M{\"u}ck}}

Until recently, configurations combining Ruby and multicore Arm systems were not properly supported.
We have revamped the \lstinline[breaklines,breakatwhitespace]|MOESI_CMP_ directory| protocol and made it the default when building gem5 for Arm.
Several issues that resulted in protocol deadlocks (especially when scaling up to many-core configurations) were fixed.
Other fixes include support for functional accesses, DMA bugs, and improved modeling of cache and directory latencies.
Additionally, support for load-linked/store-conditional (LL/SC) operations was added to the \lstinline|MESI_Three_Level| protocol, which enables it to be used with Arm as well.

\subsection[Garnet Network Model]{Garnet Network Model\footnote{By Srikant Bharadwaj and Tushar Krishna}}
\label{sec:garnet}

The interconnection system within gem5 is modeled in various levels of detail and provides extensive
flexibility to model a variety of modern systems.
The interconnect models are present within the cache-coherent Ruby memory system of gem5 (currently, Garnet cannot be used with the classic caches).
It provides the ability to create arbitrary network topologies including both homogeneous and heterogeneous systems.

There are two major variants of network models available within the Ruby memory system today: simple and Garnet.
The simple network models the routers, links, and the latencies involved with low fidelity.
This is appropriate for simulations that can sacrifice detailed interconnection network modeling for faster simulation.
The Garnet model adds detailed router microarchitecture with cycle-level buffering, resource contention, and flow control mechanisms~\cite{garnet-2}.
This model is suitable for studies that focus on interconnection units and data flow patterns.

Currently, gem5 implements an upgraded Garnet 2.0 model which provides custom routing algorithms, routers and links that support heterogeneous latencies, and standalone network simulation support.
These features allow detailed studies of on-chip networks as well as support for highly flexible topologies.
Garnet is moving to version 3.0 with the release of HeteroGarnet.
HeteroGarnet will improve Garnet by supporting modern heterogeneous systems such as 2.5D integration systems, MCM based architectures, and futuristic interconnect designs such as optical networks~\cite{kite}.
We are also working to include support for recent work on routerless NoCs~\cite{AlazemiABC18, LinPPC20}.

\subsection[GPU Compute Model]{GPU Compute Model\footnote{by Anthony Gutierrez}}
\label{sec:gpu}

GPUs have become an important part of the system design for high-performance
computing, machine learning, and many other workloads. Thus, we have integrated
a compute-based GPU model into gem5~\cite{GutierrezBeckmann2018-amdAPU}. The
GPU model is based on the AMD Graphics Core Next (GCN) architecture
\footnote{http://amd.com/en/technologies/gcn} and currently
supports the AMD GCN3 ISA.

While the initial release of the GPU model supports the GCN architecture and
GCN3 ISA, the design provides enough flexibility to implement any
modern GPU architecture that follows a single-instruction, multiple-data
(SIMD) execution model. This is achieved by following the gem5
convention of hiding ISA- and architecture-specific functionality behind a
generic API. By following this convention, the GPU core models
can be made generic enough to be used across a variety of GPU ISAs, similar
to the CPU models.

\subsubsection[GPU Thread Context]{GPU Thread Context}
At a high level the GPU core's execution model can be described as a
single-instruction, multiple-thread (SIMT) execution model. In
this execution model threads (i.e., SIMD lanes) are grouped together into
a \textit{wavefront}
(using OpenCL terminology, also called a warp in CUDA terminology) and execute the
same instruction in lock-step on a single SIMD unit.

The wavefront class in gem5 includes all the relevant GPU thread state and
executes on a single SIMD unit within a compute unit (CU). Multiple wavefronts
can be combined to form a \textit{work-group} (also called
a thread block in CUDA terminology) and multiple work-groups can be combined to
form the GPU kernel. While wavefronts map to single SIMD units, a work-group maps
to a CU.

\subsubsection[Compute Unit Pipeline]{Compute Unit Pipeline}
The CU is the primary processing unit in the GPU, which contains many
CUs. The current model in gem5 supports in-order issue of SIMD and scalar
instructions.
The workhorse of the compute unit is the SIMD unit, which executes vector
ALU instructions. The CU model is highly configurable, allowing users to set the
number of SIMD units per CU as well as the width of each SIMD unit. In addition to
the SIMD units, each CU can have some arbitrary number of scalar units, scratch
pad memory, and global memory units (both vector and scalar memory).

The SIMD unit relies on large vector register files (VRF), however the GPU can
also be configured to include a scalar register file (SRF) for use with scalar
execution units. Additionally, vector ALU and memory instructions may contain
scalar operands, and thus vector operations require access to the SRF.

At a high level the CU pipeline modeled in gem5 consists of four stages: \textit{fetch},
\textit{scoreboard check}, \textit{schedule}, and \textit{execute}. The fetch stage
is responsible for arbitrating instruction fetch, storing fetched data,
and decoding instructions for all wavefronts on the CU. The scoreboard check stage is
responsible for ensuring that intra-wavefront dependences are
correctly handled before instructions can be scheduled. The schedule stage is
responsible for arbitrating among the wavefronts presented by
the scoreboard check stage for each execution unit type. Once an instruction is dispatched
to an execution unit it is guaranteed to execute.
This ``execute at execute'' approach is the same as used by the gem5 in-order and
out-of-order CPU models. The execute stage executes the instruction by calling the \textit{execute}
method on the instruction. It manages which resources are used and dictates the timing of
operations and register file reads and writes.

\subsubsection[Kernel Launch Flow]{Kernel Launch Flow}
Kernel launch is initiated by a host application making the appropriate kernel
launch call provided by the GPU's runtime library, for example the gem5 GPU model
currently supports the ROCm software stack for GPU compute. The kernel launch
call places a kernel launch packet (i.e., a descriptor of the kernel) in memory
and notifies the GPU hardware of
the kernel launch. The GPU contains a command processor (CP) that is responsible
for handling these calls from the host as well as scheduling kernels to the GPU.

When the host launches a kernel, the packet processor in gem5 receives a signal
indicating that a kernel launch packet is ready, extracts the packet from memory,
then reads important metadata from the packet in order to begin dispatching
work and initializing the kernel's state. After the launch packet has been parsed,
information about the kernel's resource requirements are sent to the dispatcher.

As previously described wavefronts are dispatched to a CU's SIMD units, while a work-group is mapped
to a single CU. Wavefronts and work-groups, however, are combined together to form an entire kernel
which is launched to the GPU. An entire work-group (and thus all wavefronts in a work-group) must be
launched together and on the same CU, however not all work-groups must be launched simultaneously.

Kernel launch is resource limited, and the dispatcher
in the GPU is responsible for tracking work-groups, both in-flight and waiting work-groups, and
dispatching them to the CUs as resources become available. In order for a work-group to be dispatched
the CU must have enough resources to support all of its wavefronts (e.g., register file space).
Once a work-group is ready for dispatch, the CP is notified and the CU resources are reserved.
Finally, the register file state for each wavefront in the work-group is initialized based
on the metadata extracted by the CP.

\subsubsection[Autonomous Data-Race-Free GPU Tester]{Autonomous Data-Race-Free GPU Tester\footnote{by Tuan Ta}}

The Ruby coherence protocol tester is designed for CPU-like memory systems that implement relatively strong memory consistency models (e.g., TSO) and hardware-based coherence protocols (e.g., MESI).
In such systems, once a processor sends a request to memory, the request appears globally to the rest of the system.
Without knowing implementation details of target memory systems, the tester can rely on the issuing order of reads and writes to determine the current state of shared memory.
However, existing GPU memory systems are often based on weaker consistency models (e.g., sequential consistency for data-race-free programs) and implement software-directed cache coherence protocols (e.g., the VIPER Ruby protocol which requires explicit cache flushes and invalidations from software to maintain cache coherence).
The order in which reads and writes appear globally can be different from the order they are issued from GPU cores.
Therefore, the previous CPU-centric Ruby tester is not applicable to testing GPU memory systems.

The gem5 simulator has added support for an autonomous random data-race-free testing framework to validate GPU memory systems.
The tester works by randomly generating and injecting sequences of data-race-free reads and writes that are synchronized by proper atomic operations and memory fences.
By maintaining the data-race freedom of all generated sequences, the tester is able to validate responses from the system under test.
The tester is also able to periodically check for forward progress of the system and report possible deadlock and livelock.
Once encountering a failure, the tester generates an event log that captures only memory transactions related to the failure, which significantly eases the debugging process.
Ta et al. show how the tester effectively detected bugs in the implementation of VIPER protocol in gem5~\cite{Ta2019gputesting}.

\subsection[Runtime Power Modeling and DVFS Support]{Runtime Power Modeling and DVFS Support\footnote{by Stephan Diestelhorst}}
\label{sec:dvfs}

Virtually all processing today needs to consider not just aspects of performance, but also that of energy and power consumption.
Many systems are constrained by power or thermal conditions (mobile devices, boosting of desktop systems) or need to operate as energy efficiently as possible (in HPC and data centers).
We have added support to gem5 to model power-relevant silicon structures: voltage and frequency domains.
We have also added a model for enabling DVFS (dynamic voltage and frequency scaling) and support devices that allow for DVFS control by operating system governors and autonomous control.
Finally, we added an activity-based power modeling framework that measures key microarchitectural events, voltage, and frequency and allows detailed aggregation of power consumed over time similar to McPAT~\cite{LiAhn2009-mcpat, LiAhn2013-mcpat}.
Spiliopoulos et al. show that gem5's DVFS support can be integrated into both Linux and Android operating systems to provide end-to-end power and energy modeling~\cite{SpiliopoulosBHAK13}.
Additionally, these model have been extended to include power consumption caused by the activity of the SVE vector units.

\subsection[Timing-agnostic models: VirtIO and NoMali]{Timing-agnostic models: VirtIO and NoMali\footnote{By Andreas Sandberg}}
\label{sec:virtio-nomali}

With the introduction of KVM support, it quickly became apparent that some of gem5's device models, such as the IDE disk interface and the UART, were not efficient in a virtualized environment.
We also realized that these devices do not provide any relevant timing information in most experimental setups.
In fact, they are not even representative of the devices found in modern computer systems.
Similarly, when simulating mobile workload, such as Android, the GPU has a large impact on system behavior.
While it is possible to simulate an Android system without a GPU (the system resorts to software rendering), such simulations are wildly inaccurate for many CPU-side metrics~\cite{nomali}.

These problems lead to the development of a new class of device timing-agnostic models in gem5.
For block devices, pass through file systems, and serial ports, we developed a set of VirtIO-based device models.
These models only provide limited memory system interactions and no timing.
To solve the software rendering issue, we introduced a NoMali stub GPU model~\cite{nomali} that exposes the same register interface as an Arm Mali T-series and early G-series of GPUs.
This makes it possible to use a full production GPU driver stack in a simulated system without simulating the actual GPU.

\subsection[dist-gem5: Support for Distributed System Modeling]{dist-gem5: Support for Distributed System Modeling\footnote{by Mohammad Alian}}
\label{sec:dist-gem5}

Designing distributed systems requires careful analysis of the complex interplay between processor
microarchitecture, memory subsystem, inter-node network, and software layers.
However, simulating a multi-node computer system using one gem5 process is very time consuming.
Responding to the need for efficient simulation of multi-node computer systems, dist-gem5 enables parallel and distributed simulation of a hierarchical compute cluster using multiple gem5 processes.
The dist-gem5 configuration script spawns several gem5 processes, in which each of them can simulate one or several computer systems (i.e., compute node) or a scale-out network topology (i.e., network node).
Then, dist-gem5 automatically launches these gem5 processes, forwards simulated packets between them through TCP connections, and performs quantum-based synchronization to ensure correct and deterministic simulation~\cite{AlianDarbaz2016-gem5Dist, AlianKim2016-pd-gem5}.

More specifically, dist-gem5 consists of the following three main components:

\textbf{Packet forwarding:} dist-gem5 establishes a TCP socket connection between each compute node and a corresponding port of the network node to (i) forward simulated packets between compute nodes
through the simulated network topology and (ii) exchange synchronization messages.
Within each gem5 process, dist-gem5 launches a receiver thread that runs in parallel with the main simulation thread to free the main simulation thread from polling on the TCP connections.

\textbf{Synchronization:} In addition to network topology simulation, the network node implements a
synchronization barrier for performing quantum-based synchronization.
The dist-gem5 framework schedules a global sync event every quantum in each gem5 process that sends out a ``sync request'' message through the TCP connection to the network node
and waits for the reception of a ``sync ack'' to start the next quantum simulation.

\textbf{Distributed checkpointing:} dist-gem5 supports distributed checkpointing by capturing the external inter-gem5 process states including the in-flight packets inside the network node.
To ensure that no in-flight message exists between gem5 processes when the distributed checkpoint is taken, dist-gem5 only initiates checkpoints at a periodic global sync event.


\subsection[SystemC Integration]{SystemC Integration}
\label{sec:systemc}

While the open and configurable architecture of gem5 is of particular interest
in academia, one of industry's main tools for virtual prototyping is SystemC
Transaction Level Modeling (TLM)~\cite{systemc_ieee11}. Many hardware vendors
provide SystemC TLM models of their IP and there are tools, such as Synopsys
Platform Architect\footnote{\url{https://www.synopsys.com/verification/virtual-prototyping/platform-architect.html}},
that assist in building a virtual system and analyzing it. Also, many research
projects use SystemC TLM, as they benefit from the rich ecosystem of accurate
off-the-shelf models of real hardware components. However, there is a lack of
accurate and modifiable CPU models in SystemC since the model providers want to
protect their IP.
Thus, we have taken steps to make gem5 and SystemC models compatible so that researchers can construct systems using models from both frameworks at the same time.

\subsubsection[gem5 to SystemC Bridge]{gem5 to SystemC Bridge\footnote{By Christian Menard, Matthias Jung, Abdul Mutaal Ahmad, and Jeronimo Castrillon}}

SystemC TLM and gem5 were developed around the same time and are based on
similar underlying ideas. As a consequence, the hardware model used by TLM is
surprisingly close to the model of gem5. In both approaches, the system is
organized as a set of components that communicate by exchanging data packets
via a well defined protocol. The protocol abstracts over the physical
connection wires that would be used in a register transfer level (RTL)
simulation and thereby significantly increases simulation speed. In gem5,
components use \emph{requestor} and \emph{responder} ports to communicate to other
components, whereas in SystemC TLM, connections are established via
\emph{initiator} and \emph{target} sockets. Also, the three protocols
\emph{atomic}, \emph{timing} and \emph{functional} provided by gem5 find their
equivalent in the \emph{blocking}, \emph{non-blocking} and \emph{debug}
protocols of TLM. The major difference in both protocols is the treatment of
backpressure, which is implemented by a retry phase in gem5 and with the
exclusion rule of TLM.

\begin{figure}
  \begin{subfigure}{0.22\linewidth}
    \centering
    \includegraphics[height=4cm]{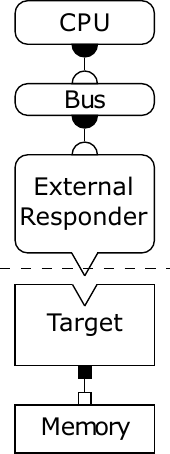}
    \caption{gem5 to SystemC}
    \label{fig:example:gem5_to_sc}
  \end{subfigure}
  ~
  \begin{subfigure}{0.22\linewidth}
    \centering
    \includegraphics[height=4cm]{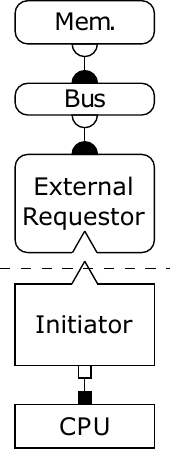}
    \caption{SystemC to gem5}
    \label{fig:example:sc_to_gem5}
  \end{subfigure}
  \begin{subfigure}{0.52\linewidth}
    \centering
    \includegraphics[height=4cm]{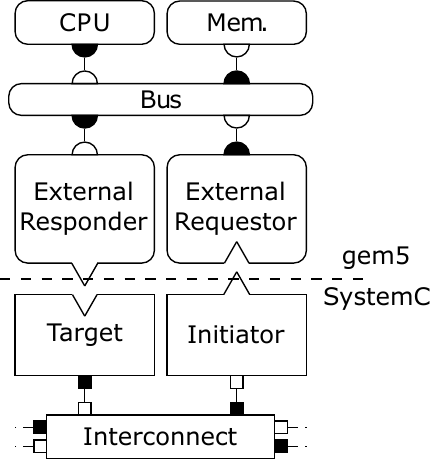}
    \caption{Both directions}
    \label{ffig:example:twoway}
  \end{subfigure}
  \caption{Possible scenarios for binding gem5 and SystemC.}
  \label{fig:gem5_tlm_example}
\end{figure}

The similarity of the two approaches enabled us to create a light-weight
compatibility layer. In our approach, co-simulation is achieved by hosting the
gem5 simulation on top of a SystemC simulation. For this, we replaced the gem5
discrete event kernel with a SystemC process that is managed by the SystemC
kernel. A set of transactors further enables communication between the two
simulation domains by translating between the two protocols as is shown in
Figure~\ref{fig:gem5_tlm_example}. Menard et al. documented our approach and showed
that the transaction between gem5 and TLM only introduces a low overhead of
about \(8\%\)~\cite{menard2017-system-systemc}.
The source code as well as basic usage examples can be found in
\texttt{util/tlm} of the gem5 repository.

\subsubsection[SystemC in gem5]{SystemC in gem5\footnote{By Gabriel Black}}

Alternatively, gem5 also has its own built in SystemC kernel and TLM implementation, and can run models natively as long as they are recompiled with gem5's SystemC header files.
These models can then use gem5's configuration mechanism and be controlled from Python, and, by using modified versions of the bridges developed to run gem5 within a SystemC simulation, TLM sockets can be connected to gem5's native ports.

This approach integrates models into gem5 more cleanly and fully since they are now first class gem5 models with access to all of gem5's APIs.
Existing models and \lstinline|c_main| implementations can generally be used as-is without any source level modifications; they just need to be recompiled against gem5's SystemC headers and linked into a gem5 binary.

While some parts of gem5's SystemC implementation are taken from the open source reference implementation (most of the data structure library and TLM), the core implementation is new and based off of the SystemC standard.
This means that code which depends on nonstandard features, behaviors, and implementation specific details of the reference implementation may not compile or work properly within gem5.
That said, gem5's SystemC kernel passes almost all of the reference implementation's test suite.
The few exceptions are tests that are broken, tests that explicitly check for implementation specific behavior, or tests for deprecated and undocumented features.

\subsection[System Call Emulation Mode Improvements]{System Call Emulation Mode Improvements\footnote{by Brandon Potter}}
\label{sec:se-mode}

System call emulation mode (SE-mode) allows gem5 to execute user-mode binaries without executing the kernel-mode system calls of a real operating system.
This is equivalent to the ``user space emulator'' in QEMU with the addition of the timing models of gem5.
Basic functionality existed in the original gem5 release~\cite{Binkert-gem5-2011}, but major improvements have been made in the past few years.
Recent additions improve the usability and increase the variety of workloads which may run in SE-mode.

\subsubsection{Dynamic Executables}

For many years, gem5 supported only statically linked executables.
This limitation prevented evaluation of workloads which require dynamic linking and loading.
To support these workloads, the SE-mode infrastructure was modified to support dynamic executables using the standard Executable and Linking Format (ELF).

At a high level, the internal ELF loader was altered to detect the need for an interpreter---the tool responsible for handling dynamic loaded libraries.
When an interpreter is required, the ELF loader will load both the interpreter and the workload into the process address space within the simulator (the guest memory in the simulator, see Figure~\ref{fig:gem5-fs-se}).
The ELF loader will also initialize stack variables to help the interpreter and the workload find each other.
With the interpreter in the address space, the workload will delegate lookups (function bindings) to the interpreter which will fixup function call invocation points on behalf of the workload.


This dynamic executable support can be combined with the virtual file system described below in Section~\ref{sec:vfs} to enable cross-platform compatibility.
With this support, users can run dynamically linked SE-mode binaries compiled for any ISA on any host ISA as long as the dynamic linker, loader, and libraries are present on the host machine.

\subsubsection{Threading Library Support}

With dynamic executable support, users encounter issues with libraries which depend on pthreads.
Many common libraries have a dependency on the pthread library.
To meet the dependency, we decided to directly support usage of native threading libraries.
To use the native threading libraries, we leverage the dynamic executable support to make standard system libraries like pthreads available to the workload.
To use this feature, the user must ensure that enough thread contexts have been allocated in their configuration script to support all threads.


The threading library support required changes to the SE-mode infrastructure.
Specifically, the clone system call required support for many new options and the futex system call required significant work.

\subsubsection{Virtual File System}
\label{sec:vfs}

In SE-mode, many system call implementations rely on functionality provided by the host machine.
For example, a workload's invocation of the ``open'' system call will cause the gem5 CPU model to hand control over to the simulator.
The SE-mode ``open'' implementation will then call the glibc open function on the host machine (which in-turn uses the host machine's open system call).
Effectively, the system call is passed from the simulated process space down to the host machine.
In Figure~\ref{fig:gem5-fs-se}, the ``Syscall Emul'' layer is implemented as a passthrough to the host operating system in many cases.

There are several reasons to employ passthrough: it avoids reimplementing complicated operating system features, it promotes code reuse by not specializing the system call implementation for each ISA, and it allows the host resources to be utilized directly from the simulated process.

However, there are several drawbacks stemming from passthrough as well.
For instance, passthrough can create API mismatches for system calls which rely on glibc library implementations. Specifically, a system call's options may differ for simulated ISA and the host ISA.
Using passthrough also can cause ABI mismatches for system calls which directly call into the host system call without interpreting system call parameters.
Additionally, passthrough can create issues when utilizing some host resources.

The virtual file system provides a solution for the third drawback, specifically, for filesystem handling.
When files are touched by the simulated process, the results of the accesses or modifications passthrough to the host filesystem.
For some cases, this causes problems.
For example, reading the contents of ``/proc/cpuinfo'' will report back results which differ from the simulated system's configuration.
In another example, the workload might try to open ``/dev/thing'' for device access which should be handled by the simulated device, not passed to the host device.

To obviate these problem, the virtual file system provides a level of indirection to catch filesystem path evaluations and modify them before the passthrough occurs.
Any path on the simulator can be redirected to any location on the host similar to mounting volumes in docker.
The \lstinline|key:value| strings for path redirection can be set via the Python configuration files.

\subsubsection{AMD ROCm v1.6}

At the time of publication, a specific version of the ROCm software stack can be used with x86 ISA builds and the GPU compute model~\ref{sec:gpu}.
The ROCm~v1.6 libraries can be loaded and used on both RHEL6 and Ubuntu 16.04 operating systems.
We distribute a set of docker containers and dockerfiles to help users get started using this specific version of ROCm with gem5.

\subsection[Testing in gem5]{Testing in gem5\footnote{by Bobby R. Bruce}}
\label{sec:testing}

In order to ensure the quality of gem5, we have continued to improve testing.
Testing ensures errors during development are caught and rectified early, prior
to release. A good testing infrastructure is essential for open-source projects
such as gem5, due frequent contributions, from many individuals, with varying
levels of expertise and familiarity with the project. Testing gives assurances
that these regular contributions are not breaking key pieces of functionality,
thus allowing for a relatively high throughput of changes.

Furthermore, due to gem5's age and size, tests give developers a degree of confidence when undertaking engineering on seldom touched component, thereby improving the productivity of all developers.
Good tests and regular testing are thereby critical in ensuring the smooth running of the project both presently and into the future.

\subsubsection{The TestLib Framework}

The majority of gem5 tests are run via \emph{TestLib}, our python-based
testing framework. TestLib runs compiled instances of gem5 on specific
computer architecture simulations, and verifies that they run as intended. It
has been designed with extendibility in-mind, allowing for tests to be easily
added as they are required. While powerful, it is important to note that
TestLib runs system-level tests, and is therefore not well-suited to testing
individual gem5 components.

\subsubsection{Unit Tests}

In order to test at a finer granularity, unit tests have been developed and
incorporated into our testing processes. Unlike our TestLib tests, these unit
tests ensure the functionality of gem5 classes and functions. Thus, if broken,
the source of the bugs can more quickly be determined. Unit Test coverage is
improving over time, and will play a greater role in improving stability in
future releases of gem5.

\subsubsection{Continuous Integration}

Via our Gerrit code-review system, we run tests for each
submitted patch prior to merging into the gem5 code base. These tests
compile gem5 against key ISA targets and run a suite of tests designed to
cover all major gem5 components (in practice, all unit tests and a subset of
the TestLib tests). This procedure supplements our code review system, thereby
reducing the possibility of new contributions introducing bugs.

In addition to the tests run prior to submission, a larger set of tests are
available for more ``in-depth'' testing. These can take several hours to complete
execution, and test functionality such as
full-system booting. We run these tests regularly to ensure gem5 meets our
standards. While these are currently triggered manually, we hope to
automatically run these tests nightly and automatically send error reports to
the developers.

\subsection{Internal gem5 Improvements and Features}
\label{sec:internal}

It is important to recognize not only all of the ground-breaking additions to the models in gem5, but also general improvements to the simulation infrastructure.
Although these improvements do not always result in new research findings, they are a key \emph{enabling factor} for the research conducted using gem5.

The simulator core of gem5 provides support for event-driven execution, statistics, and many other important functions.
These parts of the simulator are some of the most stable components, and, as part of the gem5-20 release and in the subsequent releases, we will be defining stable APIs for these interfaces.
By making these interfaces \emph{stable} APIs, it will facilitate long-term support for integrating other simulators (e.g., SystemC as discussed in Section~\ref{sec:systemc} and SST) and projects that build off of gem5 (e.g., gem5-gpu~\cite{PowerHOHW15}, gem5-aladdin~\cite{ShaoXSWB16}, gem5-graphics~\cite{GubranAamodt2019-emerald}, and many others.)

\subsubsection[HDF5 Support]{HDF5 Support\footnote{by Andreas Sandberg}}

A major change in the latest gem5 release is the new statistics API.
While the driver for this API was to improve support for hierarchical statistics formats like HDF5~\cite{hdf5}, there are other more tangible benefits as well.
Unlike the old API where all statistics live in the same namespace, the new API introduces a notion of statistics groups.
In most typical use cases, statistics are bound to the current SimObject's group, which is then bound to its parent by the runtime.
This ensures that there is a tree of statistics groups that match the SimObject graph.
However, groups are not limited to SimObject.
Behind the scenes, this reduces the amount of boiler plate code when defining statistics and makes the code far less error prone.
The new API also brings benefits to simulation scripts.
A feature many users have requested in the past has been the ability to dump statistics for a subset of the object graph.
This is now possible by passing a SimObject to the stat dump call, which limits the statistics dump to that subtree of the graph.

With the new statistics API in place, it became possible to support hierarchical data formats like HDF5.
Unlike gem5's traditional text-based statistics files, HDF5 stores data in a binary file format that resembles a file system.
Unlike the traditional text files, HDF5 has a rich ecosystem of tools and official bindings for many popular languages, including Python and R.
In addition to making analysis easier, the HDF5 backend is optimized for storing time series data.
HDF5 files internally store data as N-dimensional matrices.
In gem5's implementation, we use one dimension for time and the remaining dimensions for the statistic we want to represent.
For example, a scalar statistic is represented as a 1-dimensional vector.
When analyzing such series using Python, the HDF5 backend imports such data sets as a standard NumPy array that can be used in common data analysis and visualization flows.
The additional data needed to support filesystem-like structures inside the stat files introduces some storage overheads.
However, these are quickly amortized when sampling statistics since the incremental storage needed for every sample is orders of magnitude smaller than the traditional text-based statistics format.

\subsubsection[Python 3]{Python 3\footnote{by Andreas Sandberg and Giacomo Travaglini}}

One of the main features which separates gem5 from other architectural simulators is its robust support for scripting.
The main interface to configuring and running gem5 simulations is Python scripts.
While the fundamental design has not changed, there have been many changes to the underlying implementation over the past years.
The original implementation frequently suffered from bugs in the code generated by SWIG and usability was hampered by poor adherence to modern standards in SWIG's C++ parser.
The move to PyBind11~\cite{pybind11} greatly improved the reliability of the bindings by removing the need for a separate C++ parser, and made it easier to expose new functionality to Python in a reliable and type-safe manner.

The migration from SWIG to PyBind11 also provided a good starting point for the more ambitious project of making gem5 Python 3 compatible, which is now complete.
This has not had a direct impact on the gem5 feature set yet, but it ensures that the simulator will continue to run on Linux distributions that are released in 2020 and onwards.
However, it does enable exciting improvements under the hood.
A couple of good examples are type annotations that can be used to enable better static code analysis and greatly improved string formatting.
Our ambition is to completely phase out Python 2 support in the near future to benefit from these new features.

\subsubsection[Asynchronous Modeling in gem5]{Asynchronous Modeling in gem5\footnote{by Giacomo Travaglini}}

The difficulties of writing a complex hardware model within gem5 is that your model needs to be able to work and be representative of the simulated hardware in both atomic and timing mode.

For simple devices which only respond to requests, this is usually not a concern.
The situation gets worse when the device can send requests and responses or has DMA capabilities.
A method generating and forwarding a read packet needs to differentiate between atomic and timing behavior by handling the first with a blocking operation (the read returns the value as soon as the forwarding method returns) and the second with a non-blocking call: the value will be returned later in time.
The situation becomes dramatic in timing mode if multiple sequential DMAs are stacked so that any read operation depends on previous ones; this is the case for page table walks for example.

This software design problem has been elegantly solved using coroutines.
Coroutines allow you to execute your task, checkpoint it, and resume it later from where you stopped.
To be more specific to our use case, you can tag your DMA packets with the coroutine itself, and you could resume the coroutine once the device receives the read response.

While waiting for coroutines to be fully supported in C++20, we have implemented a coroutine library within gem5 that allows developers to use coroutines to generate asynchronous models.
The coroutine class is built on top of a ``Fiber'' class, which was a pre-existing symmetric coroutine implementation, and it provides boost-like APIs to the user.

At the moment coroutines are used by the SMMUv3 model developed and the GICv3 ITS model (Interrupt Translation Service).
There are many other use cases for this API in other gem5 models, and we are planning on updating those models in the future.

\subsection[Updating Guest<->Simulator APIs]{Updating Guest$\leftrightarrow$Simulator APIs\footnote{By Gabriel Black}}
\label{sec:guest-sim}

It is sometimes helpful or necessary for gem5 to interact with the software running inside the simulation in some non-architectural way.
In Figure~\ref{fig:gem5-fs-fs}, the application under test may want to call a function \emph{in the gem5 simulator} or vice versa.
For instance, gem5 might want to intervene and adjust the guest's behavior to skip over some uninteresting function, like one that sets all of physical memory to zeroes, or which uses a loop to measure CPU speed or implement a delay.
It might also want to monitor guest behavior to know when something important like a kernel panic has happened.
Guest software might also want to purposefully request some behavior from gem5 such as requesting that gem5 exit, recording the current value of the simulation statistics, taking a checkpoint,  and reading or writing a file on the host, etc.

One way the simulator can react to guest behavior is by executing a callback when the guest executes a certain program counter (PC).
The PC would generally come from the table of symbols loaded with, for instance, an OS kernel, and would let gem5 detect when certain kernel functions were about to execute.
This mechanism has been improved to make it easier for different types of CPU models to implement.
These include the CPU models which use KVM and the Arm Fast Model based CPUs.

The gem5$\leftrightarrow$guest interaction might also be triggered by the application running on the guest itself.
One common way to use these mechanisms from within the guest is to use the ``m5'' utility which parses command line arguments and then triggers whatever gem5 behavior was requested.
This utility is in the process of being revamped so that support is consistent across ISAs, along with many other improvements including supporting all the back end mechanisms described above.

Because it is not possible to universally predict what PCs correspond to requests from the guest, a different signaling mechanism is necessary.
Traditional gem5 CPU models redefined unused opcodes from the target ISA for that purpose.
However, this mechanism is not universal.
For instance, when using the KVM-based CPU model instructions behave like they would on real hardware since they are running on real hardware.
In these special cases, we require other APIs.

Finally, the gem5 simulator code must be able to decipher the calling convention of guest code. Historically this was done in several different ways.
These were somewhat redundant, inconsistent, incomplete, and difficult to maintain.

We have implemented a new system of templates to pull apart a function's signature and marshal arguments from within the guest automatically.
Those arguments are then used to call an arbitrary function in gem5.
Once the function finishes, it can optionally return a value into the guest if it wants to override or just observe guest behavior.

For instance, suppose we had the function shown in Figure~\ref{fig:code1}.
If we wanted to call it from within the guest using calling convention AAPCS32, once gem5 had detected the call (as described above), it could call \lstinline|foo()| with arguments from the guest as shown in Figure~\ref{fig:code2}.

\begin{figure}
    \centering
    \begin{subfigure}{\linewidth}
        \begin{lstlisting}[frame=single,basicstyle=\small]
int
foo(char bar, float baz)
{
    return (baz < 0) ? bar : bar + 1;
}
        \end{lstlisting}
        \caption{Example guest$\leftrightarrow$function. \vspace{10pt}}
        \label{fig:code1}
    \end{subfigure}

    \begin{subfigure}{\linewidth}
        \begin{lstlisting}[frame=single,basicstyle=\small]
invokeSimcall<Aapcs32>(tc, foo);
        \end{lstlisting}
        \caption{Example gem5 code.}
        \label{fig:code2}
    \end{subfigure}
    \caption{Example use of new Guest$\leftrightarrow$Simulator APIs}
\end{figure}









\section{Conclusion}

Over the past nine years, the gem5 simulator has become an increasingly important tool in the computer architecture research community.
This paper describes the significant strides taken to improve this community-developed infrastructure.
Looking forward, with the continued support of the broader computer architecture research community, the gem5 simulator will continue to mature and its use will continue to grow.
The community will continue to add new features, add new models, and increase the stability of the simulator.

The overarching goal of the future development of the gem5 simulator is to increase its user base by expanding its use both within the computer architecture community and in other computer systems research fields.
To accomplish this goal, we will be providing ``known-good'' configurations and other tools to enable reproducible computer system simulation.
We will also provide more user support to broaden the gem5 community through improved documentation and learning materials.
Through these efforts, we look forward to continue to grow and improve the gem5 simulation infrastructure through the next 20 years of computer system development.

\section{Acknowledgements}
\label{sec:acks}

The development of gem5 is community-driven and distributed.
The contributions to the gem5 community go beyond just the source code, and many people who have contributed to the broader gem5 community are not acknowledged here.

We would like to specially acknowledge the late Nathan Binkert.
Nate was a driving force behind the creation of gem5 and without his vision and his dedication to code quality this open-source community infrastructure would not be the success that it is today.

The gem5 project management committee consists of Bradford Beckmann, Gabriel Black, Anthony Gutierrez, Jason Lowe-Power (chair), Steven Reinhardt, Ali Saidi, Andreas Sandberg, Matthew Sinclair, Giacomo Travaglini, and David Wood.
Previous members include Nathan Binkert, and Andreas Hansson.
The project management committee manages the administration of the project and ensures that the gem5 community runs smoothly.

This work is supported in part by the National Science Foundation (CNS-1925724, CNS-1925485, CNS-1850566, and many others) and Brookhaven National Laboratory.
Google has donated resources to host gem5's codes, code review, continuous integration, and other web-based resources.

This work was partially completed with funding from the European Union's Horizon 2020 research and innovation programme under project Mont-Blanc 2020, grant agreement 779877.

We would also like to thank all of the other contributors to gem5 including
Chris Adeniyi-Jones, Michael Adler, Neha Agarwal, John Alsop, Lluc Alvarez, Ricardo Alves, Matteo Andreozzi, Ruben Ayrapetyan, Erfan Azarkhish, Akash Bagdia, Jairo Balart, Marco Balboni, Marc Mari Barcelo, Andrew Bardsley, Isaac S\'anchez Barrera, Maurice Becker, Rizwana Begum, Glenn Bergmans, Umesh Bhaskar, Nathan Binkert, Sascha Bischoff, Geoffrey Blake, Maximilien Breughe, Kevin Brodsky, Ruslan Bukin, Pau Cabre, Javier Cano-Cano, Emilio Castillo, Jiajie Chen, James Clarkson, Stan Czerniawski, Stanislaw Czerniawski, Sandipan Das, Nayan Deshmukh, Cagdas Dirik, Xiangyu Dong, Gabor Dozsa, Ron Dreslinski, Curtis Dunham, Alexandru Dutu, Yasuko Eckert, Sherif Elhabbal, Hussein Elnawawy, Marco Elver, Chris Emmons, Fernando Endo, Peter Enns, Matt Evans, Mbou Eyole, Matteo M. Fusi, Giacomo Gabrielli, Santi Galan, Victor Garcia, Mrinmoy Ghosh, Pritha Ghoshal, Riken Gohil, Rekai Gonzalez-Alberquilla, Brian Grayson, Samuel Grayson, Edmund Grimley-Evans, Thomas Grocutt, Joe Gross, David Guillen-Fandos, Deyaun Guo, Anders Handler, David Hashe, Mitch Hayenga, Blake Hechtman, Javier Bueno Hedo, Eric Van Hensbergen, Joel Hestness, Mark Hildebrand, Matthias Hille, Rune Holm, Chun-Chen Hsu, Lisa Hsu, Hsuan Hsu, Stian Hvatum, Tom Jablin, Nuwan Jayasena, Min Kyu Jeong, Ola Jeppsson, Jakub Jermar, Sudhanshu Jha, Ian Jiang, Dylan Johnson, Daniel Johnson, Rene de Jong, John Kalamatianos, Khalique, Do\u{g}ukan Korkmazt\"urk, Georg Kotheimer, Djordje Kovacevic, Robert Kovacsics, Rohit Kurup, Anouk Van Laer, Jan-Peter Larsson, Michael LeBeane, Jui-min Lee, Michael Levenhagen, Weiping Liao, Pin-Yen Lin, Nicholas Lindsay, Yifei Liu, Gabe Loh, Andrew Lukefahr, Palle Lyckegaard, Jiuyue Ma, Xiaoyu Ma, Andriani Mappoura, Jose Marinho, Bertrand Marquis, Maxime Martinasso, Sean McGoogan, Mingyuan, Monir Mozumder, Malek Musleh, Earl Ou, Xin Ouyang, Rutuja Oza, Andrea Pellegrini, Arthur Perais, Adrien Pesle, Polydoros Petrakis, Anis Peysieux, Christoph Pfister, Sujay Phadke, Ivan Pizarro, Matthew Poremba, Brandon Potter, Siddhesh Poyarekar, Nathanael Premillieu, Sooraj Puthoor, Jing Qu, Prakash Ramrakhyani, Steve Reinhardt, Isaac Richter, Paul Rosenfeld, Shawn Rosti, Ali Saidi, Karthik Sangaiah, Ciro Santilli, Robert Scheffel, Sophiane Senni, Korey Sewell, Faissal Sleiman, Maximilian Stein, Po-Hao Su, Chander Sudanthi, Kanishk Sugand, Dam Sunwoo, Koan-Sin Tan, Michiel Van Tol, Erik Tomusk, Christopher Torng, Ashkan Tousi, Sergei Trofimov, Avishai Tvila, Ani Udipi, Jordi Vaquero, Llu\'is Vilanova, Wade Walker, Yu-hsin Wang, Bradley Wang, William Wang, Moyang Wang, Zicong Wang, Vince Weaver, Uri Wiener, Sean Wilson, Severin Wischmann, Willy Wolff, Yuan Yao, Jieming Yin, Bjoern A. Zeeb, Dongxue Zhang, Tao Zhang, Xianwei Zhang, Zhang Zheng, Chuan Zhu, Chen Zou.

\copyright 2020 Advanced Micro Devices, Inc. All rights reserved.
AMD, the AMD Arrow logo, and combinations thereof are trademarks of Advanced
Micro Devices, Inc. Other product names used in this publication are for
identification purposes only and may be trademarks of their respective
companies.

\bibliographystyle{ACM-Reference-Format}
\bibliography{references}


\begin{thebibliography}{77}


\ifx \showCODEN    \undefined \def \showCODEN     #1{\unskip}     \fi
\ifx \showDOI      \undefined \def \showDOI       #1{#1}\fi
\ifx \showISBNx    \undefined \def \showISBNx     #1{\unskip}     \fi
\ifx \showISBNxiii \undefined \def \showISBNxiii  #1{\unskip}     \fi
\ifx \showISSN     \undefined \def \showISSN      #1{\unskip}     \fi
\ifx \showLCCN     \undefined \def \showLCCN      #1{\unskip}     \fi
\ifx \shownote     \undefined \def \shownote      #1{#1}          \fi
\ifx \showarticletitle \undefined \def \showarticletitle #1{#1}   \fi
\ifx \showURL      \undefined \def \showURL       {\relax}        \fi
\providecommand\bibfield[2]{#2}
\providecommand\bibinfo[2]{#2}
\providecommand\natexlab[1]{#1}
\providecommand\showeprint[2][]{arXiv:#2}

\bibitem[\protect\citeauthoryear{??}{sys}{2012}]%
        {systemc_ieee11}
 \bibinfo{year}{2012}\natexlab{}.
\newblock \showarticletitle{{IEEE} Standard for Standard {SystemC} Language
  Reference Manual}.
\newblock \bibinfo{journal}{\emph{IEEE Std 1666-2011 (Revision of IEEE Std
  1666-2005)}} (\bibinfo{date}{Jan} \bibinfo{year}{2012}).
\newblock
\urldef\tempurl%
\url{https://doi.org/10.1109/IEEESTD.2012.6134619}
\showDOI{\tempurl}


\bibitem[\protect\citeauthoryear{Agarwal, Krishna, Peh, and Jha}{Agarwal
  et~al\mbox{.}}{2009}]%
        {garnet-2}
\bibfield{author}{\bibinfo{person}{Niket Agarwal}, \bibinfo{person}{Tushar
  Krishna}, \bibinfo{person}{Li-Shiuan Peh}, {and} \bibinfo{person}{Niraj~K
  Jha}.} \bibinfo{year}{2009}\natexlab{}.
\newblock \showarticletitle{GARNET: A detailed on-chip network model inside a
  full-system simulator}. In \bibinfo{booktitle}{\emph{IEEE International
  Symposium on Performance Analysis of Systems and Software, 2009. ISPASS
  2009.}} IEEE, \bibinfo{pages}{33--42}.
\newblock


\bibitem[\protect\citeauthoryear{Akram and Sawalha}{Akram and Sawalha}{2016}]%
        {akram201686}
\bibfield{author}{\bibinfo{person}{Ayaz Akram} {and} \bibinfo{person}{Lina
  Sawalha}.} \bibinfo{year}{2016}\natexlab{}.
\newblock \showarticletitle{{x86 Computer Architecture Simulators: A
  Comparative Study}}. In \bibinfo{booktitle}{\emph{{IEEE 34th International
  Conference on Computer Design}}} \emph{(\bibinfo{series}{ICCD})}. IEEE,
  \bibinfo{pages}{638--645}.
\newblock


\bibitem[\protect\citeauthoryear{Akram and Sawalha}{Akram and Sawalha}{2019}]%
        {akram2019validation}
\bibfield{author}{\bibinfo{person}{Ayaz Akram} {and} \bibinfo{person}{Lina
  Sawalha}.} \bibinfo{year}{2019}\natexlab{}.
\newblock \showarticletitle{Validation of the gem5 Simulator for x86
  Architectures}. In \bibinfo{booktitle}{\emph{2019 IEEE/ACM Performance
  Modeling, Benchmarking and Simulation of High Performance Computer Systems
  (PMBS)}}. IEEE, \bibinfo{pages}{53--58}.
\newblock


\bibitem[\protect\citeauthoryear{Alameldeen and Agarwal}{Alameldeen and
  Agarwal}{2018}]%
        {alameldeen2018opportunistic}
\bibfield{author}{\bibinfo{person}{Alaa~R Alameldeen} {and}
  \bibinfo{person}{Rajat Agarwal}.} \bibinfo{year}{2018}\natexlab{}.
\newblock \showarticletitle{Opportunistic compression for direct-mapped DRAM
  caches}. In \bibinfo{booktitle}{\emph{Proceedings of the International
  Symposium on Memory Systems}}. ACM, \bibinfo{pages}{129--136}.
\newblock


\bibitem[\protect\citeauthoryear{Alazemi, AziziMazreah, Bose, and Chen}{Alazemi
  et~al\mbox{.}}{2018}]%
        {AlazemiABC18}
\bibfield{author}{\bibinfo{person}{Fawaz Alazemi}, \bibinfo{person}{Arash
  AziziMazreah}, \bibinfo{person}{Bella Bose}, {and} \bibinfo{person}{Lizhong
  Chen}.} \bibinfo{year}{2018}\natexlab{}.
\newblock \showarticletitle{Routerless Network-on-Chip}. In
  \bibinfo{booktitle}{\emph{{IEEE} International Symposium on High Performance
  Computer Architecture, {HPCA} 2018, Vienna, Austria, February 24-28, 2018}}.
  \bibinfo{publisher}{{IEEE} Computer Society}, \bibinfo{pages}{492--503}.
\newblock
\urldef\tempurl%
\url{https://doi.org/10.1109/HPCA.2018.00049}
\showDOI{\tempurl}


\bibitem[\protect\citeauthoryear{Alian, Darbaz, Dozsa, Diestelhorst, Kim, and
  Kim}{Alian et~al\mbox{.}}{2017}]%
        {AlianDarbaz2016-gem5Dist}
\bibfield{author}{\bibinfo{person}{M. Alian}, \bibinfo{person}{U. Darbaz},
  \bibinfo{person}{G. Dozsa}, \bibinfo{person}{S. Diestelhorst},
  \bibinfo{person}{D. Kim}, {and} \bibinfo{person}{N.~S. Kim}.}
  \bibinfo{year}{2017}\natexlab{}.
\newblock \showarticletitle{{dist-gem5: Distributed Simulation of Computer
  Clusters}}. In \bibinfo{booktitle}{\emph{{IEEE International Symposium on
  Performance Analysis of Systems and Software}}}
  \emph{(\bibinfo{series}{ISPASS})}. \bibinfo{pages}{153--162}.
\newblock
\urldef\tempurl%
\url{https://doi.org/10.1109/ISPASS.2017.7975287}
\showDOI{\tempurl}


\bibitem[\protect\citeauthoryear{Alian, Kim, and Kim}{Alian
  et~al\mbox{.}}{2016}]%
        {AlianKim2016-pd-gem5}
\bibfield{author}{\bibinfo{person}{M. Alian}, \bibinfo{person}{D. Kim}, {and}
  \bibinfo{person}{N.~Sung Kim}.} \bibinfo{year}{2016}\natexlab{}.
\newblock \showarticletitle{{pd-gem5: Simulation Infrastructure for
  Parallel/Distributed Computer Systems}}.
\newblock \bibinfo{journal}{\emph{{IEEE Computer Architecture Letters}}}
  \bibinfo{number}{01} (\bibinfo{date}{jan} \bibinfo{year}{2016}),
  \bibinfo{pages}{41--44}.
\newblock
\showISSN{1556-6064}
\urldef\tempurl%
\url{https://doi.org/10.1109/LCA.2015.2438295}
\showDOI{\tempurl}


\bibitem[\protect\citeauthoryear{Alves and Felton}{Alves and Felton}{2004}]%
        {ArmTustZone}
\bibfield{author}{\bibinfo{person}{Tiago Alves} {and} \bibinfo{person}{Don
  Felton}.} \bibinfo{year}{2004}\natexlab{}.
\newblock \showarticletitle{{TrustZone: Integrated Hardware and Software
  Security}}.
\newblock \bibinfo{journal}{\emph{Information Quarterly}}
  (\bibinfo{year}{2004}), \bibinfo{pages}{18--24}.
\newblock


\bibitem[\protect\citeauthoryear{{AMD}}{{AMD}}{2012}]%
        {gcnWhitepaper}
\bibfield{author}{\bibinfo{person}{{AMD}}.} \bibinfo{year}{2012}\natexlab{}.
\newblock \bibinfo{title}{{AMD Graphics Core Next (GCN) Architecture}}.
\newblock
  \bibinfo{howpublished}{\url{https://www.techpowerup.com/gpu-specs/docs/amd-gcn1-architecture.pdf}}.
\newblock


\bibitem[\protect\citeauthoryear{{AMD}}{{AMD}}{2016}]%
        {gcn3Manual}
\bibfield{author}{\bibinfo{person}{{AMD}}.} \bibinfo{year}{2016}\natexlab{}.
\newblock \bibinfo{title}{{Graphics Core Next Architecture, Generation 3}}.
\newblock
  \bibinfo{howpublished}{\url{http://developer.amd.com/wordpress/media/2013/12/AMD_GCN3_Instruction_Set_Architecture_rev1.1.pdf}}.
\newblock


\bibitem[\protect\citeauthoryear{Arm Ltd.}{Arm Ltd.}{2020}]%
        {ArmARM}
Arm Ltd. \bibinfo{year}{2020}\natexlab{}.
\newblock \bibinfo{booktitle}{\emph{Arm\textregistered Architecture Reference
  Manual: Armv8, for Armv8-A architecture profile} (\bibinfo{edition}{f.b}
  ed.)}.
\newblock Arm Ltd.
\newblock
\urldef\tempurl%
\url{https://developer.arm.com/docs/ddi0487/fb/arm-architecture-reference-manual-armv8-for-armv8-a-architecture-profile}
\showURL{%
\tempurl}


\bibitem[\protect\citeauthoryear{Asri, Pedram, John, and Gerstlauer}{Asri
  et~al\mbox{.}}{2016}]%
        {asri2016simulator}
\bibfield{author}{\bibinfo{person}{Mochamad Asri}, \bibinfo{person}{Ardavan
  Pedram}, \bibinfo{person}{Lizy~K John}, {and} \bibinfo{person}{Andreas
  Gerstlauer}.} \bibinfo{year}{2016}\natexlab{}.
\newblock \showarticletitle{{Simulator Calibration for Accelerator-Rich
  Architecture Studies}}. In \bibinfo{booktitle}{\emph{International Conference
  on Embedded Computer Systems: Architectures, Modeling and Simulation
  (SAMOS),}}. IEEE, \bibinfo{pages}{88--95}.
\newblock


\bibitem[\protect\citeauthoryear{Bailey, Barszcz, Barton, Browning, Carter,
  Dagum, Fatoohi, Frederickson, Lasinski, Schreiber, et~al\mbox{.}}{Bailey
  et~al\mbox{.}}{1991}]%
        {npb}
\bibfield{author}{\bibinfo{person}{David~H Bailey}, \bibinfo{person}{Eric
  Barszcz}, \bibinfo{person}{John~T Barton}, \bibinfo{person}{David~S
  Browning}, \bibinfo{person}{Robert~L Carter}, \bibinfo{person}{Leonardo
  Dagum}, \bibinfo{person}{Rod~A Fatoohi}, \bibinfo{person}{Paul~O
  Frederickson}, \bibinfo{person}{Thomas~A Lasinski}, \bibinfo{person}{Rob~S
  Schreiber}, {et~al\mbox{.}}} \bibinfo{year}{1991}\natexlab{}.
\newblock \showarticletitle{The NAS parallel benchmarks}.
\newblock \bibinfo{journal}{\emph{The International Journal of Supercomputing
  Applications}} \bibinfo{volume}{5}, \bibinfo{number}{3}
  (\bibinfo{year}{1991}), \bibinfo{pages}{63--73}.
\newblock


\bibitem[\protect\citeauthoryear{Beamer, Asanovi{\'c}, and Patterson}{Beamer
  et~al\mbox{.}}{2015}]%
        {gapbs}
\bibfield{author}{\bibinfo{person}{Scott Beamer}, \bibinfo{person}{Krste
  Asanovi{\'c}}, {and} \bibinfo{person}{David Patterson}.}
  \bibinfo{year}{2015}\natexlab{}.
\newblock \showarticletitle{The GAP benchmark suite}.
\newblock \bibinfo{journal}{\emph{arXiv preprint arXiv:1508.03619}}
  (\bibinfo{year}{2015}).
\newblock


\bibitem[\protect\citeauthoryear{{Bharadwaj}, {Yin}, {Beckmann}, and
  {Krishna}}{{Bharadwaj} et~al\mbox{.}}{2020}]%
        {kite}
\bibfield{author}{\bibinfo{person}{S. {Bharadwaj}}, \bibinfo{person}{J. {Yin}},
  \bibinfo{person}{B. {Beckmann}}, {and} \bibinfo{person}{T. {Krishna}}.}
  \bibinfo{year}{2020}\natexlab{}.
\newblock \showarticletitle{Kite: A Family of Heterogeneous Interposer
  Topologies Enabled via Accurate Interconnect Modeling}. In
  \bibinfo{booktitle}{\emph{2020 57th ACM/IEEE Design Automation Conference
  (DAC)}}.
\newblock


\bibitem[\protect\citeauthoryear{Bienia, Kumar, Singh, and Li}{Bienia
  et~al\mbox{.}}{2008}]%
        {parsec}
\bibfield{author}{\bibinfo{person}{Christian Bienia}, \bibinfo{person}{Sanjeev
  Kumar}, \bibinfo{person}{Jaswinder~Pal Singh}, {and} \bibinfo{person}{Kai
  Li}.} \bibinfo{year}{2008}\natexlab{}.
\newblock \showarticletitle{The PARSEC benchmark suite: Characterization and
  architectural implications}. In \bibinfo{booktitle}{\emph{Proceedings of the
  17th International Conference on Parallel Architectures and Compilation
  techniques}}. \bibinfo{pages}{72--81}.
\newblock


\bibitem[\protect\citeauthoryear{Binkert, Beckmann, Black, Reinhardt, Saidi,
  Basu, Hestness, Hower, Krishna, Sardashti, Sen, Sewell, Shoaib, Vaish, Hill,
  and Wood}{Binkert et~al\mbox{.}}{2011}]%
        {Binkert-gem5-2011}
\bibfield{author}{\bibinfo{person}{Nathan Binkert}, \bibinfo{person}{Bradford
  Beckmann}, \bibinfo{person}{Gabriel Black}, \bibinfo{person}{Steven~K.
  Reinhardt}, \bibinfo{person}{Ali Saidi}, \bibinfo{person}{Arkaprava Basu},
  \bibinfo{person}{Joel Hestness}, \bibinfo{person}{Derek~R. Hower},
  \bibinfo{person}{Tushar Krishna}, \bibinfo{person}{Somayeh Sardashti},
  \bibinfo{person}{Rathijit Sen}, \bibinfo{person}{Korey Sewell},
  \bibinfo{person}{Muhammad Shoaib}, \bibinfo{person}{Nilay Vaish},
  \bibinfo{person}{Mark~D. Hill}, {and} \bibinfo{person}{David~A. Wood}.}
  \bibinfo{year}{2011}\natexlab{}.
\newblock \showarticletitle{The {g}em5 Simulator}.
\newblock \bibinfo{journal}{\emph{SIGARCH Compututer Architecture News}}
  \bibinfo{volume}{39}, \bibinfo{number}{2} (\bibinfo{date}{Aug.}
  \bibinfo{year}{2011}), \bibinfo{pages}{1--7}.
\newblock
\showISSN{0163-5964}
\urldef\tempurl%
\url{https://doi.org/10.1145/2024716.2024718}
\showDOI{\tempurl}


\bibitem[\protect\citeauthoryear{Binkert, Dreslinski, Hsu, Lim, Saidi, and
  Reinhardt}{Binkert et~al\mbox{.}}{2006}]%
        {BinkertDHLSR06}
\bibfield{author}{\bibinfo{person}{Nathan~L. Binkert},
  \bibinfo{person}{Ronald~G. Dreslinski}, \bibinfo{person}{Lisa~R. Hsu},
  \bibinfo{person}{Kevin~T. Lim}, \bibinfo{person}{Ali~G. Saidi}, {and}
  \bibinfo{person}{Steven~K. Reinhardt}.} \bibinfo{year}{2006}\natexlab{}.
\newblock \showarticletitle{The {M5} Simulator: Modeling Networked Systems}.
\newblock \bibinfo{journal}{\emph{{IEEE} Micro}} \bibinfo{volume}{26},
  \bibinfo{number}{4} (\bibinfo{year}{2006}), \bibinfo{pages}{52--60}.
\newblock
\urldef\tempurl%
\url{https://doi.org/10.1109/MM.2006.82}
\showDOI{\tempurl}


\bibitem[\protect\citeauthoryear{Bucek, Lange, and von Kistowski}{Bucek
  et~al\mbox{.}}{2018}]%
        {spec17}
\bibfield{author}{\bibinfo{person}{James Bucek}, \bibinfo{person}{Klaus-Dieter
  Lange}, {and} \bibinfo{person}{Jóakim von Kistowski}.}
  \bibinfo{year}{2018}\natexlab{}.
\newblock \showarticletitle{SPEC CPU2017: Next-Generation Compute Benchmark}.
\newblock \bibinfo{journal}{\emph{ICPE '18: Companion of the 2018 ACM/SPEC
  International Conference on Performance Engineering}},
  \bibinfo{pages}{41--42}.
\newblock
\showISBNx{978-1-4503-5629-9}
\urldef\tempurl%
\url{https://doi.org/10.1145/3185768.3185771}
\showDOI{\tempurl}


\bibitem[\protect\citeauthoryear{Butko, Garibotti, Ost, and Sassatelli}{Butko
  et~al\mbox{.}}{2012}]%
        {butko2012accuracy}
\bibfield{author}{\bibinfo{person}{Anastasiia Butko}, \bibinfo{person}{Rafael
  Garibotti}, \bibinfo{person}{Luciano Ost}, {and} \bibinfo{person}{Gilles
  Sassatelli}.} \bibinfo{year}{2012}\natexlab{}.
\newblock \showarticletitle{{Accuracy Evaluation of GEM5 Simulator System}}. In
  \bibinfo{booktitle}{\emph{IEEE 7th International Workshop on Reconfigurable
  Communication-centric Systems-on-Chip}}. \bibinfo{publisher}{York, UK},
  \bibinfo{pages}{1--7}.
\newblock


\bibitem[\protect\citeauthoryear{Chandrasekar, Weis, Li, Akesson, Naji, Jung,
  Wehn, and Goossens}{Chandrasekar et~al\mbox{.}}{2014}]%
        {kargoo_14}
\bibfield{author}{\bibinfo{person}{Karthik Chandrasekar},
  \bibinfo{person}{Christian Weis}, \bibinfo{person}{Yonghui Li},
  \bibinfo{person}{Benny Akesson}, \bibinfo{person}{Omar Naji},
  \bibinfo{person}{Matthias Jung}, \bibinfo{person}{Norbert Wehn}, {and}
  \bibinfo{person}{Kees Goossens}.} \bibinfo{year}{2014}\natexlab{}.
\newblock \bibinfo{title}{{DRAMP}ower: {O}pen-source {DRAM} power \& energy
  estimation tool}.
\newblock \bibinfo{howpublished}{\url{ http://www.drampower.info}}.
\newblock


\bibitem[\protect\citeauthoryear{Chen, Yang, Dick, Shang, and Lekatsas}{Chen
  et~al\mbox{.}}{2010}]%
        {chen2010c}
\bibfield{author}{\bibinfo{person}{Xi Chen}, \bibinfo{person}{Lei Yang},
  \bibinfo{person}{Robert~P Dick}, \bibinfo{person}{Li Shang}, {and}
  \bibinfo{person}{Haris Lekatsas}.} \bibinfo{year}{2010}\natexlab{}.
\newblock \showarticletitle{C-pack: A high-performance microprocessor cache
  compression algorithm}.
\newblock \bibinfo{journal}{\emph{Very Large Scale Integration (VLSI) Systems,
  IEEE Transactions on}} \bibinfo{volume}{18}, \bibinfo{number}{8}
  (\bibinfo{year}{2010}), \bibinfo{pages}{1196--1208}.
\newblock


\bibitem[\protect\citeauthoryear{Conway, Kalyanasundharam, Donley, Lepak, and
  Hughes}{Conway et~al\mbox{.}}{2010}]%
        {Conway:opteron:2010}
\bibfield{author}{\bibinfo{person}{Pat Conway}, \bibinfo{person}{Nathan
  Kalyanasundharam}, \bibinfo{person}{Gregg Donley}, \bibinfo{person}{Kevin
  Lepak}, {and} \bibinfo{person}{Bill Hughes}.}
  \bibinfo{year}{2010}\natexlab{}.
\newblock \showarticletitle{Cache Hierarchy and Memory Subsystem of the {AMD}
  Opteron Processor}.
\newblock \bibinfo{journal}{\emph{{IEEE} Micro}} \bibinfo{volume}{30},
  \bibinfo{number}{2} (\bibinfo{year}{2010}), \bibinfo{pages}{16--29}.
\newblock
\urldef\tempurl%
\url{https://doi.org/10.1109/MM.2010.31}
\showDOI{\tempurl}


\bibitem[\protect\citeauthoryear{{de Jong} and {Sandberg}}{{de Jong} and
  {Sandberg}}{2016}]%
        {nomali}
\bibfield{author}{\bibinfo{person}{R. {de Jong}} {and} \bibinfo{person}{A.
  {Sandberg}}.} \bibinfo{year}{2016}\natexlab{}.
\newblock \showarticletitle{NoMali: Simulating a realistic graphics driver
  stack using a stub GPU}. In \bibinfo{booktitle}{\emph{2016 IEEE International
  Symposium on Performance Analysis of Systems and Software (ISPASS)}}.
  \bibinfo{pages}{255--262}.
\newblock


\bibitem[\protect\citeauthoryear{Dutu and Slice}{Dutu and Slice}{2015}]%
        {DutuSlice2015-kvm}
\bibfield{author}{\bibinfo{person}{Alexandru Dutu} {and} \bibinfo{person}{John
  Slice}.} \bibinfo{year}{2015}\natexlab{}.
\newblock \showarticletitle{{KVM CPU Model in Syscall Emulation Mode}}. In
  \bibinfo{booktitle}{\emph{{Second gem5 User Workshop}}}.
\newblock


\bibitem[\protect\citeauthoryear{Endo, Courouss{\'e}, and Charles}{Endo
  et~al\mbox{.}}{2014}]%
        {endo2014micro}
\bibfield{author}{\bibinfo{person}{Fernando~A Endo}, \bibinfo{person}{Damien
  Courouss{\'e}}, {and} \bibinfo{person}{Henri-Pierre Charles}.}
  \bibinfo{year}{2014}\natexlab{}.
\newblock \showarticletitle{{Micro-architectural Simulation of In-Order and
  Out-of-Order ARM Microprocessors with gem5}}. In
  \bibinfo{booktitle}{\emph{International Conference on Embedded Computer
  Systems: Architectures, Modeling, and Simulation (SAMOS XIV), 2014}}. IEEE,
  \bibinfo{pages}{266--273}.
\newblock


\bibitem[\protect\citeauthoryear{Gubran and Aamodt}{Gubran and Aamodt}{2019}]%
        {GubranAamodt2019-emerald}
\bibfield{author}{\bibinfo{person}{Ayub~A. Gubran} {and}
  \bibinfo{person}{Tor~M. Aamodt}.} \bibinfo{year}{2019}\natexlab{}.
\newblock \showarticletitle{{Emerald: Graphics Modeling for SoC Systems}}. In
  \bibinfo{booktitle}{\emph{{Proceedings of the 46th International Symposium on
  Computer Architecture}}} (Phoenix, Arizona) \emph{(\bibinfo{series}{ISCA
  '19})}. \bibinfo{publisher}{Association for Computing Machinery},
  \bibinfo{address}{New York, NY, USA}, \bibinfo{pages}{169--182}.
\newblock
\showISBNx{9781450366694}
\urldef\tempurl%
\url{https://doi.org/10.1145/3307650.3322221}
\showDOI{\tempurl}


\bibitem[\protect\citeauthoryear{Gutierrez, Beckmann, Dutu, Gross, LeBeane,
  Kalamatianos, Kayiran, Poremba, Potter, Puthoor, Sinclair, Wyse, Yin, Zhang,
  Jain, and Rogers}{Gutierrez et~al\mbox{.}}{2018}]%
        {GutierrezBeckmann2018-amdAPU}
\bibfield{author}{\bibinfo{person}{Anthony Gutierrez},
  \bibinfo{person}{Bradford~M. Beckmann}, \bibinfo{person}{Alexandru Dutu},
  \bibinfo{person}{Joseph Gross}, \bibinfo{person}{Michael LeBeane},
  \bibinfo{person}{John Kalamatianos}, \bibinfo{person}{Onur Kayiran},
  \bibinfo{person}{Matthew Poremba}, \bibinfo{person}{Brandon Potter},
  \bibinfo{person}{Sooraj Puthoor}, \bibinfo{person}{Matthew~D. Sinclair},
  \bibinfo{person}{Michael Wyse}, \bibinfo{person}{Jieming Yin},
  \bibinfo{person}{Xianwei Zhang}, \bibinfo{person}{Akshay Jain}, {and}
  \bibinfo{person}{Timothy Rogers}.} \bibinfo{year}{2018}\natexlab{}.
\newblock \showarticletitle{{Lost in Abstraction: Pitfalls of Analyzing GPUs at
  the Intermediate Language Level}}. In \bibinfo{booktitle}{\emph{{Proceedings
  of 24th IEEE International Symposium on High Performance Computer
  Architecture}}} \emph{(\bibinfo{series}{HPCA})}. \bibinfo{pages}{608--619}.
\newblock
\showISSN{2378-203X}
\urldef\tempurl%
\url{https://doi.org/10.1109/HPCA.2018.00058}
\showDOI{\tempurl}


\bibitem[\protect\citeauthoryear{Gutierrez, Pusdesris, Dreslinski, Mudge,
  Sudanthi, Emmons, Hayenga, and Paver}{Gutierrez et~al\mbox{.}}{2014}]%
        {gutierrez2014sources}
\bibfield{author}{\bibinfo{person}{Anthony Gutierrez}, \bibinfo{person}{Joseph
  Pusdesris}, \bibinfo{person}{Ronald~G Dreslinski}, \bibinfo{person}{Trevor
  Mudge}, \bibinfo{person}{Chander Sudanthi}, \bibinfo{person}{Christopher~D
  Emmons}, \bibinfo{person}{Mitchell Hayenga}, {and} \bibinfo{person}{Nigel
  Paver}.} \bibinfo{year}{2014}\natexlab{}.
\newblock \showarticletitle{{Sources of Error in Full-System Simulation}}. In
  \bibinfo{booktitle}{\emph{IEEE International Symposium on Performance
  Analysis of Systems and Software}}. \bibinfo{publisher}{Monterey, CA},
  \bibinfo{pages}{13--22}.
\newblock


\bibitem[\protect\citeauthoryear{{Hansson}, {Agarwal}, {Kolli}, {Wenisch}, and
  {Udipi}}{{Hansson} et~al\mbox{.}}{2014}]%
        {HanssonAgarwal2014-gem5DRAM}
\bibfield{author}{\bibinfo{person}{A. {Hansson}}, \bibinfo{person}{N.
  {Agarwal}}, \bibinfo{person}{A. {Kolli}}, \bibinfo{person}{T. {Wenisch}},
  {and} \bibinfo{person}{A.~N. {Udipi}}.} \bibinfo{year}{2014}\natexlab{}.
\newblock \showarticletitle{{Simulating DRAM controllers for future system
  architecture exploration}}. In \bibinfo{booktitle}{\emph{{IEEE International
  Symposium on Performance Analysis of Systems and Software}}}
  \emph{(\bibinfo{series}{ISPASS})}. \bibinfo{pages}{201--210}.
\newblock


\bibitem[\protect\citeauthoryear{Hennessy and Patterson}{Hennessy and
  Patterson}{2018}]%
        {HennessyPatterson-turingLect-isca18}
\bibfield{author}{\bibinfo{person}{John Hennessy} {and} \bibinfo{person}{David
  Patterson}.} \bibinfo{year}{2018}\natexlab{}.
\newblock \bibinfo{title}{{A New Golden Age for Computer Architecture: Domain
  Specific Hardware/Software Co-Design, Enhanced Security, Open Instruction
  Sets, and Agile Chip Development}}.
\newblock \bibinfo{howpublished}{{Turing Award Lecture}}.
\newblock


\bibitem[\protect\citeauthoryear{Hennessy and Patterson}{Hennessy and
  Patterson}{2019}]%
        {HennessyPatterson-CACM19}
\bibfield{author}{\bibinfo{person}{John~L. Hennessy} {and}
  \bibinfo{person}{David~A. Patterson}.} \bibinfo{year}{2019}\natexlab{}.
\newblock \showarticletitle{A New Golden Age for Computer Architecture}.
\newblock \bibinfo{journal}{\emph{Commun. ACM}} \bibinfo{volume}{62},
  \bibinfo{number}{2} (\bibinfo{date}{Jan.} \bibinfo{year}{2019}),
  \bibinfo{pages}{48--60}.
\newblock
\showISSN{0001-0782}
\urldef\tempurl%
\url{https://doi.org/10.1145/3282307}
\showDOI{\tempurl}


\bibitem[\protect\citeauthoryear{Henning}{Henning}{2006}]%
        {spec06}
\bibfield{author}{\bibinfo{person}{John~L Henning}.}
  \bibinfo{year}{2006}\natexlab{}.
\newblock \showarticletitle{SPEC CPU2006 benchmark descriptions}.
\newblock \bibinfo{journal}{\emph{ACM SIGARCH Computer Architecture News}}
  \bibinfo{volume}{34}, \bibinfo{number}{4} (\bibinfo{year}{2006}),
  \bibinfo{pages}{1--17}.
\newblock


\bibitem[\protect\citeauthoryear{Hsieh, Pedretti, Meng, Coskun, Levenhagen, and
  Rodrigues}{Hsieh et~al\mbox{.}}{2012}]%
        {HsiehPedretti2012-sst-gem5}
\bibfield{author}{\bibinfo{person}{Mingyu Hsieh}, \bibinfo{person}{Kevin
  Pedretti}, \bibinfo{person}{Jie Meng}, \bibinfo{person}{Ayse Coskun},
  \bibinfo{person}{Michael Levenhagen}, {and} \bibinfo{person}{Arun
  Rodrigues}.} \bibinfo{year}{2012}\natexlab{}.
\newblock \showarticletitle{{SST + gem5 = a Scalable Simulation Infrastructure
  for High Performance Computing}}. In \bibinfo{booktitle}{\emph{{Proceedings
  of the 5th International ICST Conference on Simulation Tools and
  Techniques}}} (Desenzano del Garda, Italy) \emph{(\bibinfo{series}{SIMUTOOLS
  ’12})}. \bibinfo{publisher}{ICST (Institute for Computer Sciences,
  Social-Informatics and Telecommunications Engineering)},
  \bibinfo{address}{Brussels, BEL}, \bibinfo{pages}{196--201}.
\newblock
\showISBNx{9781450315104}


\bibitem[\protect\citeauthoryear{Jagtap, Diestelhorst, Hansson, Jung, and
  Wehn}{Jagtap et~al\mbox{.}}{2016}]%
        {jagdie_16}
\bibfield{author}{\bibinfo{person}{Radhika Jagtap}, \bibinfo{person}{Stephan
  Diestelhorst}, \bibinfo{person}{Andreas Hansson}, \bibinfo{person}{Matthias
  Jung}, {and} \bibinfo{person}{Norbert Wehn}.}
  \bibinfo{year}{2016}\natexlab{}.
\newblock \showarticletitle{{E}xploring {S}ystem {P}erformance using {E}lastic
  {T}races: {F}ast, {A}ccurate and {P}ortable}. In
  \bibinfo{booktitle}{\emph{IEEE International Conference on Embedded Computer
  Systems Architectures Modeling and Simulation (SAMOS), July, 2016, Samos
  Island, Greece}}.
\newblock


\bibitem[\protect\citeauthoryear{Jagtap, Jung, Elsasser, Weis, Hansson, and
  Wehn}{Jagtap et~al\mbox{.}}{2017}]%
        {jagjun_17}
\bibfield{author}{\bibinfo{person}{Radhika Jagtap}, \bibinfo{person}{Matthias
  Jung}, \bibinfo{person}{Wendy Elsasser}, \bibinfo{person}{Christian Weis},
  \bibinfo{person}{Andreas Hansson}, {and} \bibinfo{person}{Norbert Wehn}.}
  \bibinfo{year}{2017}\natexlab{}.
\newblock \showarticletitle{{I}ntegrating {DRAM} {P}ower-{D}own {M}odes in gem5
  and {Q}uantifying their {I}mpact}. In \bibinfo{booktitle}{\emph{International
  Symposium on Memory Systems (MEMSYS17)}}.
\newblock


\bibitem[\protect\citeauthoryear{Jakob, Rhinelander, and Moldovan}{Jakob
  et~al\mbox{.}}{2017}]%
        {pybind11}
\bibfield{author}{\bibinfo{person}{Wenzel Jakob}, \bibinfo{person}{Jason
  Rhinelander}, {and} \bibinfo{person}{Dean Moldovan}.}
  \bibinfo{year}{2017}\natexlab{}.
\newblock \bibinfo{title}{pybind11 -- Seamless operability between C++11 and
  Python}.
\newblock
\newblock
\newblock
\shownote{https://github.com/pybind/pybind11.}


\bibitem[\protect\citeauthoryear{Jaleel, Theobald, Steely, and Emer}{Jaleel
  et~al\mbox{.}}{2010}]%
        {Jaleel2010rrip}
\bibfield{author}{\bibinfo{person}{Aamer Jaleel}, \bibinfo{person}{Kevin~B.
  Theobald}, \bibinfo{person}{Simon~C. Steely}, {and} \bibinfo{person}{Joel
  Emer}.} \bibinfo{year}{2010}\natexlab{}.
\newblock \showarticletitle{High Performance Cache Replacement Using
  Re-Reference Interval Prediction (RRIP)}. In
  \bibinfo{booktitle}{\emph{Proceedings of the 37th Annual International
  Symposium on Computer Architecture}} (Saint-Malo, France)
  \emph{(\bibinfo{series}{ISCA '10})}. \bibinfo{publisher}{Association for
  Computing Machinery}, \bibinfo{address}{New York, NY, USA},
  \bibinfo{pages}{60--71}.
\newblock
\showISBNx{9781450300537}
\urldef\tempurl%
\url{https://doi.org/10.1145/1815961.1815971}
\showDOI{\tempurl}


\bibitem[\protect\citeauthoryear{Jo, Lee, Jang, Lee, Ajdari, and Kim}{Jo
  et~al\mbox{.}}{2018}]%
        {jo2018diagsim}
\bibfield{author}{\bibinfo{person}{Jae-Eon Jo}, \bibinfo{person}{Gyu-Hyeon
  Lee}, \bibinfo{person}{Hanhwi Jang}, \bibinfo{person}{Jaewon Lee},
  \bibinfo{person}{Mohammadamin Ajdari}, {and} \bibinfo{person}{Jangwoo Kim}.}
  \bibinfo{year}{2018}\natexlab{}.
\newblock \showarticletitle{DiagSim: Systematically Diagnosing Simulators for
  Healthy Simulations}.
\newblock \bibinfo{journal}{\emph{ACM Transactions on Architecture and Code
  Optimization (TACO)}} \bibinfo{volume}{15}, \bibinfo{number}{1}
  (\bibinfo{year}{2018}), \bibinfo{pages}{4}.
\newblock


\bibitem[\protect\citeauthoryear{Jouppi}{Jouppi}{1993}]%
        {10.1145/173682.165154}
\bibfield{author}{\bibinfo{person}{Norman~P. Jouppi}.}
  \bibinfo{year}{1993}\natexlab{}.
\newblock \showarticletitle{Cache Write Policies and Performance}.
\newblock \bibinfo{journal}{\emph{SIGARCH Compututer Architecture News}}
  \bibinfo{volume}{21}, \bibinfo{number}{2} (\bibinfo{date}{May}
  \bibinfo{year}{1993}), \bibinfo{pages}{191--201}.
\newblock
\showISSN{0163-5964}
\urldef\tempurl%
\url{https://doi.org/10.1145/173682.165154}
\showDOI{\tempurl}


\bibitem[\protect\citeauthoryear{Jung, Weis, and Wehn}{Jung
  et~al\mbox{.}}{2015}]%
        {junwei_15}
\bibfield{author}{\bibinfo{person}{Matthias Jung}, \bibinfo{person}{Christian
  Weis}, {and} \bibinfo{person}{Norbert Wehn}.}
  \bibinfo{year}{2015}\natexlab{}.
\newblock \showarticletitle{{DRAMS}ys: {A} flexible {DRAM} {S}ubsystem {D}esign
  {S}pace {E}xploration {F}ramework}.
\newblock \bibinfo{journal}{\emph{IPSJ Transactions on System LSI Design
  Methodology (T-SLDM)}} (\bibinfo{date}{August} \bibinfo{year}{2015}).
\newblock
\urldef\tempurl%
\url{https://doi.org/10.2197/ipsjtsldm.8.63}
\showDOI{\tempurl}


\bibitem[\protect\citeauthoryear{Jung, Weis, Wehn, and Chandrasekar}{Jung
  et~al\mbox{.}}{2013}]%
        {junwei_13}
\bibfield{author}{\bibinfo{person}{Matthias Jung}, \bibinfo{person}{Christian
  Weis}, \bibinfo{person}{Norbert Wehn}, {and} \bibinfo{person}{Karthik
  Chandrasekar}.} \bibinfo{year}{2013}\natexlab{}.
\newblock \showarticletitle{{TLM} modelling of 3{D} stacked wide {I}/{O} {DRAM}
  subsystems: a virtual platform for memory controller design space
  exploration}. In \bibinfo{booktitle}{\emph{Proceedings of the 2013 Workshop
  on Rapid Simulation and Performance Evaluation: Methods and Tools}} (Berlin,
  Germany) \emph{(\bibinfo{series}{RAPIDO '13})}. \bibinfo{publisher}{ACM},
  \bibinfo{address}{New York, NY, USA}, Article \bibinfo{articleno}{5},
  \bibinfo{numpages}{6}~pages.
\newblock
\showISBNx{978-1-4503-1539-5}
\urldef\tempurl%
\url{https://doi.org/10.1145/2432516.2432521}
\showDOI{\tempurl}


\bibitem[\protect\citeauthoryear{Kim, Yang, and Mutlu}{Kim
  et~al\mbox{.}}{2016}]%
        {yoongy_16}
\bibfield{author}{\bibinfo{person}{Yoongu Kim}, \bibinfo{person}{Weikun Yang},
  {and} \bibinfo{person}{Onur Mutlu}.} \bibinfo{year}{2016}\natexlab{}.
\newblock \showarticletitle{Ramulator: A Fast and Extensible DRAM Simulator}.
\newblock \bibinfo{journal}{\emph{IEEE Comput. Archit. Lett.}}
  \bibinfo{volume}{15}, \bibinfo{number}{1} (\bibinfo{date}{Jan.}
  \bibinfo{year}{2016}), \bibinfo{pages}{45--49}.
\newblock
\showISSN{1556-6056}
\urldef\tempurl%
\url{https://doi.org/10.1109/LCA.2015.2414456}
\showDOI{\tempurl}


\bibitem[\protect\citeauthoryear{Kodama, Odajima, Asato, and Sato}{Kodama
  et~al\mbox{.}}{2019}]%
        {Kodama:Riken:2019}
\bibfield{author}{\bibinfo{person}{Yuetsu Kodama}, \bibinfo{person}{Tetsuya
  Odajima}, \bibinfo{person}{Akira Asato}, {and} \bibinfo{person}{Mitsuhisa
  Sato}.} \bibinfo{year}{2019}\natexlab{}.
\newblock \showarticletitle{Evaluation of the {RIKEN} Post-K Processor
  Simulator}.
\newblock \bibinfo{journal}{\emph{CoRR}}  \bibinfo{volume}{abs/1904.06451}
  (\bibinfo{year}{2019}).
\newblock
\showeprint[arxiv]{1904.06451}
\urldef\tempurl%
\url{http://arxiv.org/abs/1904.06451}
\showURL{%
\tempurl}


\bibitem[\protect\citeauthoryear{Li, Ahn, Strong, Brockman, Tullsen, and
  Jouppi}{Li et~al\mbox{.}}{2009}]%
        {LiAhn2009-mcpat}
\bibfield{author}{\bibinfo{person}{Sheng Li}, \bibinfo{person}{Jung-Ho Ahn},
  \bibinfo{person}{Richard~D. Strong}, \bibinfo{person}{Jay~B. Brockman},
  \bibinfo{person}{Dean~M. Tullsen}, {and} \bibinfo{person}{Norman~P. Jouppi}.}
  \bibinfo{year}{2009}\natexlab{}.
\newblock \showarticletitle{{McPAT: An Integrated Power, Area, and Timing
  Modeling Framework for Multicore and Manycore Architectures}}. In
  \bibinfo{booktitle}{\emph{{42nd Annual IEEE/ACM International Symposium on
  Microarchitecture}}} \emph{(\bibinfo{series}{MICRO})}.
  \bibinfo{pages}{469--480}.
\newblock
\showISSN{1072-4451}


\bibitem[\protect\citeauthoryear{Li, Ahn, Strong, Brockman, Tullsen, and
  Jouppi}{Li et~al\mbox{.}}{2013}]%
        {LiAhn2013-mcpat}
\bibfield{author}{\bibinfo{person}{Sheng Li}, \bibinfo{person}{Jung~Ho Ahn},
  \bibinfo{person}{Richard~D. Strong}, \bibinfo{person}{Jay~B. Brockman},
  \bibinfo{person}{Dean~M. Tullsen}, {and} \bibinfo{person}{Norman~P. Jouppi}.}
  \bibinfo{year}{2013}\natexlab{}.
\newblock \showarticletitle{{The McPAT Framework for Multicore and Manycore
  Architectures: Simultaneously Modeling Power, Area, and Timing}}.
\newblock \bibinfo{journal}{\emph{{ACM Transactions on Architecture \& Code
  Optimization}}} \bibinfo{volume}{10}, \bibinfo{number}{1}, Article
  \bibinfo{articleno}{5} (\bibinfo{date}{April} \bibinfo{year}{2013}),
  \bibinfo{numpages}{29}~pages.
\newblock
\showISSN{1544-3566}
\urldef\tempurl%
\url{https://doi.org/10.1145/2445572.2445577}
\showDOI{\tempurl}


\bibitem[\protect\citeauthoryear{{Li}, {Yang}, {Reddy}, {Srivastava}, and
  {Jacob}}{{Li} et~al\mbox{.}}{2020}]%
        {dramsim3}
\bibfield{author}{\bibinfo{person}{S. {Li}}, \bibinfo{person}{Z. {Yang}},
  \bibinfo{person}{D. {Reddy}}, \bibinfo{person}{A. {Srivastava}}, {and}
  \bibinfo{person}{B. {Jacob}}.} \bibinfo{year}{2020}\natexlab{}.
\newblock \showarticletitle{DRAMsim3: a Cycle-accurate, Thermal-Capable DRAM
  Simulator}.
\newblock \bibinfo{journal}{\emph{IEEE Computer Architecture Letters}}
  (\bibinfo{year}{2020}), \bibinfo{pages}{1--1}.
\newblock


\bibitem[\protect\citeauthoryear{Lin, Penney, Pedram, and Chen}{Lin
  et~al\mbox{.}}{2020}]%
        {LinPPC20}
\bibfield{author}{\bibinfo{person}{Ting{-}Ru Lin}, \bibinfo{person}{Drew
  Penney}, \bibinfo{person}{Massoud Pedram}, {and} \bibinfo{person}{Lizhong
  Chen}.} \bibinfo{year}{2020}\natexlab{}.
\newblock \showarticletitle{A Deep Reinforcement Learning Framework for
  Architectural Exploration: {A} Routerless NoC Case Study}. In
  \bibinfo{booktitle}{\emph{{IEEE} International Symposium on High Performance
  Computer Architecture, {HPCA} 2020, San Diego, CA, USA, February 22-26,
  2020}}. \bibinfo{publisher}{{IEEE}}, \bibinfo{pages}{99--110}.
\newblock
\urldef\tempurl%
\url{https://doi.org/10.1109/HPCA47549.2020.00018}
\showDOI{\tempurl}


\bibitem[\protect\citeauthoryear{Lowe-Power}{Lowe-Power}{2015}]%
        {Power-gem5horrors-2015}
\bibfield{author}{\bibinfo{person}{Jason Lowe-Power}.}
  \bibinfo{year}{2015}\natexlab{}.
\newblock \showarticletitle{gem5 Horrors and what we can do about it}. In
  \bibinfo{booktitle}{\emph{Second gem5 User Workshop with ISCA 2015}}.
\newblock


\bibitem[\protect\citeauthoryear{Martin, Hill, and Wood}{Martin
  et~al\mbox{.}}{2003}]%
        {MartinHill2003-tokenCoh}
\bibfield{author}{\bibinfo{person}{Milo M.~K. Martin}, \bibinfo{person}{Mark~D.
  Hill}, {and} \bibinfo{person}{David~A. Wood}.}
  \bibinfo{year}{2003}\natexlab{}.
\newblock \showarticletitle{{Token Coherence: Decoupling Performance and
  Correctness}}. In \bibinfo{booktitle}{\emph{{Proceedings of the 30th Annual
  International Symposium on Computer Architecture}}} (San Diego, California)
  \emph{(\bibinfo{series}{ISCA '03})}. \bibinfo{publisher}{Association for
  Computing Machinery}, \bibinfo{address}{New York, NY, USA},
  \bibinfo{pages}{182--193}.
\newblock
\showISBNx{0769519458}
\urldef\tempurl%
\url{https://doi.org/10.1145/859618.859640}
\showDOI{\tempurl}


\bibitem[\protect\citeauthoryear{Martin, Sorin, Beckmann, Marty, Xu,
  Alameldeen, Moore, Hill, and Wood}{Martin et~al\mbox{.}}{2005}]%
        {MartinSBMXAMHW05}
\bibfield{author}{\bibinfo{person}{Milo M.~K. Martin},
  \bibinfo{person}{Daniel~J. Sorin}, \bibinfo{person}{Bradford~M. Beckmann},
  \bibinfo{person}{Michael~R. Marty}, \bibinfo{person}{Min Xu},
  \bibinfo{person}{Alaa~R. Alameldeen}, \bibinfo{person}{Kevin~E. Moore},
  \bibinfo{person}{Mark~D. Hill}, {and} \bibinfo{person}{David~A. Wood}.}
  \bibinfo{year}{2005}\natexlab{}.
\newblock \showarticletitle{Multifacet's general execution-driven
  multiprocessor simulator {(GEMS)} toolset}.
\newblock \bibinfo{journal}{\emph{{SIGARCH} Computer Architecture News}}
  \bibinfo{volume}{33}, \bibinfo{number}{4} (\bibinfo{year}{2005}),
  \bibinfo{pages}{92--99}.
\newblock
\urldef\tempurl%
\url{https://doi.org/10.1145/1105734.1105747}
\showDOI{\tempurl}


\bibitem[\protect\citeauthoryear{Menard, Castrillon, Jung, and Wehn}{Menard
  et~al\mbox{.}}{2017}]%
        {menard2017-system-systemc}
\bibfield{author}{\bibinfo{person}{Christian Menard}, \bibinfo{person}{Jeronimo
  Castrillon}, \bibinfo{person}{Matthias Jung}, {and} \bibinfo{person}{Norbert
  Wehn}.} \bibinfo{year}{2017}\natexlab{}.
\newblock \showarticletitle{System Simulation with gem5 and {SystemC}: The
  Keystone for Full Interoperability}. In \bibinfo{booktitle}{\emph{2017
  International Conference on Embedded Computer Systems: Architectures,
  Modeling, and Simulation (SAMOS)}}. \bibinfo{pages}{62--69}.
\newblock


\bibitem[\protect\citeauthoryear{Nagarajan, Sorin, Hill, and Wood}{Nagarajan
  et~al\mbox{.}}{2020}]%
        {NagarajanSorin2020-cohMCMPrimer}
\bibfield{author}{\bibinfo{person}{Vijay Nagarajan}, \bibinfo{person}{Daniel~J.
  Sorin}, \bibinfo{person}{Mark~D. Hill}, {and} \bibinfo{person}{David~A.
  Wood}.} \bibinfo{year}{2020}\natexlab{}.
\newblock \showarticletitle{{A Primer on Memory Consistency and Cache
  Coherence}}.
\newblock \bibinfo{journal}{\emph{{Synthesis Lectures on Computer
  Architecture}}} \bibinfo{volume}{15}, \bibinfo{number}{1}
  (\bibinfo{date}{February} \bibinfo{year}{2020}), \bibinfo{pages}{1--294}.
\newblock
\urldef\tempurl%
\url{https://doi.org/10.2200/S00962ED2V01Y201910CAC049}
\showDOI{\tempurl}


\bibitem[\protect\citeauthoryear{Nikoleris, Eeckhout, Hagersten, and
  Carlson}{Nikoleris et~al\mbox{.}}{2019}]%
        {NikolerisEHC19}
\bibfield{author}{\bibinfo{person}{Nikos Nikoleris}, \bibinfo{person}{Lieven
  Eeckhout}, \bibinfo{person}{Erik Hagersten}, {and} \bibinfo{person}{Trevor~E.
  Carlson}.} \bibinfo{year}{2019}\natexlab{}.
\newblock \showarticletitle{Directed Statistical Warming through Time
  Traveling}. In \bibinfo{booktitle}{\emph{Proceedings of the 52nd Annual
  {IEEE/ACM} International Symposium on Microarchitecture, {MICRO} 2019,
  Columbus, OH, USA, October 12-16, 2019}}. \bibinfo{publisher}{{ACM}},
  \bibinfo{pages}{1037--1049}.
\newblock
\urldef\tempurl%
\url{https://doi.org/10.1145/3352460.3358264}
\showDOI{\tempurl}


\bibitem[\protect\citeauthoryear{Nikoleris, Sandberg, Hagersten, and
  Carlson}{Nikoleris et~al\mbox{.}}{2016}]%
        {NikolerisSHC16}
\bibfield{author}{\bibinfo{person}{Nikos Nikoleris}, \bibinfo{person}{Andreas
  Sandberg}, \bibinfo{person}{Erik Hagersten}, {and} \bibinfo{person}{Trevor~E.
  Carlson}.} \bibinfo{year}{2016}\natexlab{}.
\newblock \showarticletitle{CoolSim: Statistical techniques to replace cache
  warming with efficient, virtualized profiling}. In
  \bibinfo{booktitle}{\emph{International Conference on Embedded Computer
  Systems: Architectures, Modeling and Simulation, {SAMOS} 2016, Agios
  Konstantinos, Samos Island, Greece, July 17-21, 2016}},
  \bibfield{editor}{\bibinfo{person}{Walid~A. Najjar} {and}
  \bibinfo{person}{Andreas Gerstlauer}} (Eds.). \bibinfo{publisher}{{IEEE}},
  \bibinfo{pages}{106--115}.
\newblock
\urldef\tempurl%
\url{https://doi.org/10.1109/SAMOS.2016.7818337}
\showDOI{\tempurl}


\bibitem[\protect\citeauthoryear{Nowatzki, Menon, Ho, and
  Sankaralingam}{Nowatzki et~al\mbox{.}}{2015}]%
        {nowatzki2015architectural}
\bibfield{author}{\bibinfo{person}{Tony Nowatzki}, \bibinfo{person}{Jaikrishnan
  Menon}, \bibinfo{person}{Chen-Han Ho}, {and} \bibinfo{person}{Karthikeyan
  Sankaralingam}.} \bibinfo{year}{2015}\natexlab{}.
\newblock \showarticletitle{Architectural simulators considered harmful}.
\newblock \bibinfo{journal}{\emph{IEEE Micro}} \bibinfo{volume}{35},
  \bibinfo{number}{6} (\bibinfo{year}{2015}), \bibinfo{pages}{4--12}.
\newblock


\bibitem[\protect\citeauthoryear{Pekhimenko, Seshadri, Mutlu, Gibbons, Kozuch,
  and Mowry}{Pekhimenko et~al\mbox{.}}{2012}]%
        {pekhimenko2012base}
\bibfield{author}{\bibinfo{person}{Gennady Pekhimenko}, \bibinfo{person}{Vivek
  Seshadri}, \bibinfo{person}{Onur Mutlu}, \bibinfo{person}{Phillip~B Gibbons},
  \bibinfo{person}{Michael~A Kozuch}, {and} \bibinfo{person}{Todd~C Mowry}.}
  \bibinfo{year}{2012}\natexlab{}.
\newblock \showarticletitle{Base-delta-immediate compression: practical data
  compression for on-chip caches}. In \bibinfo{booktitle}{\emph{Proceedings of
  the 21st International Conference on Parallel Architectures and Compilation
  Techniques}}. ACM, \bibinfo{pages}{377--388}.
\newblock


\bibitem[\protect\citeauthoryear{Petrogalli}{Petrogalli}{2020}]%
        {white-paper-on-SVE-and-VLA-programming}
\bibfield{author}{\bibinfo{person}{Francesco Petrogalli}.}
  \bibinfo{year}{2020}\natexlab{}.
\newblock \bibinfo{booktitle}{\emph{A sneak peek into SVE and VLA
  programming}}.
\newblock \bibinfo{type}{{T}echnical {R}eport}. \bibinfo{institution}{Arm Ltd.}
\newblock


\bibitem[\protect\citeauthoryear{Power, Basu, Gu, Puthoor, Beckmann, Hill,
  Reinhardt, and Wood}{Power et~al\mbox{.}}{2013}]%
        {PowerBasu2013-hsc}
\bibfield{author}{\bibinfo{person}{Jason Power}, \bibinfo{person}{Arkaprava
  Basu}, \bibinfo{person}{Junli Gu}, \bibinfo{person}{Sooraj Puthoor},
  \bibinfo{person}{Bradford~M. Beckmann}, \bibinfo{person}{Mark~D. Hill},
  \bibinfo{person}{Steven~K. Reinhardt}, {and} \bibinfo{person}{David~A.
  Wood}.} \bibinfo{year}{2013}\natexlab{}.
\newblock \showarticletitle{Heterogeneous System Coherence for Integrated
  {CPU}-{GPU} Systems}. In \bibinfo{booktitle}{\emph{Proceedings of the 46th
  Annual IEEE/ACM International Symposium on Microarchitecture}} (Davis,
  California) \emph{(\bibinfo{series}{MICRO-46})}. \bibinfo{publisher}{ACM},
  \bibinfo{address}{New York, NY, USA}, \bibinfo{pages}{457--467}.
\newblock
\showISBNx{978-1-4503-2638-4}
\urldef\tempurl%
\url{https://doi.org/10.1145/2540708.2540747}
\showDOI{\tempurl}


\bibitem[\protect\citeauthoryear{Power, Hestness, Orr, Hill, and Wood}{Power
  et~al\mbox{.}}{2015}]%
        {PowerHOHW15}
\bibfield{author}{\bibinfo{person}{Jason Power}, \bibinfo{person}{Joel
  Hestness}, \bibinfo{person}{Marc~S. Orr}, \bibinfo{person}{Mark~D. Hill},
  {and} \bibinfo{person}{David~A. Wood}.} \bibinfo{year}{2015}\natexlab{}.
\newblock \showarticletitle{gem5-gpu: {A} Heterogeneous {CPU-GPU} Simulator}.
\newblock \bibinfo{journal}{\emph{{IEEE} Comput. Archit. Lett.}}
  \bibinfo{volume}{14}, \bibinfo{number}{1} (\bibinfo{year}{2015}),
  \bibinfo{pages}{34--36}.
\newblock
\urldef\tempurl%
\url{https://doi.org/10.1109/LCA.2014.2299539}
\showDOI{\tempurl}


\bibitem[\protect\citeauthoryear{Rodrigues, Hemmert, Barrett, Kersey, Oldfield,
  Weston, Risen, Cook, Rosenfeld, Cooper-Balis, and Jacob}{Rodrigues
  et~al\mbox{.}}{2011}]%
        {RodriguesHemmert2011-sst}
\bibfield{author}{\bibinfo{person}{A.~F. Rodrigues}, \bibinfo{person}{K.~S.
  Hemmert}, \bibinfo{person}{B.~W. Barrett}, \bibinfo{person}{C. Kersey},
  \bibinfo{person}{R. Oldfield}, \bibinfo{person}{M. Weston},
  \bibinfo{person}{R. Risen}, \bibinfo{person}{J. Cook}, \bibinfo{person}{P.
  Rosenfeld}, \bibinfo{person}{E. Cooper-Balis}, {and} \bibinfo{person}{B.
  Jacob}.} \bibinfo{year}{2011}\natexlab{}.
\newblock \showarticletitle{{The Structural Simulation Toolkit}}.
\newblock \bibinfo{journal}{\emph{SIGMETRICS Perform. Eval. Rev.}}
  \bibinfo{volume}{38}, \bibinfo{number}{4} (\bibinfo{date}{March}
  \bibinfo{year}{2011}), \bibinfo{pages}{37--42}.
\newblock
\showISSN{0163-5999}
\urldef\tempurl%
\url{https://doi.org/10.1145/1964218.1964225}
\showDOI{\tempurl}


\bibitem[\protect\citeauthoryear{Roelke and Stan}{Roelke and Stan}{2017}]%
        {risc5-gem5}
\bibfield{author}{\bibinfo{person}{Alec Roelke} {and}
  \bibinfo{person}{Mircea~R. Stan}.} \bibinfo{year}{2017}\natexlab{}.
\newblock \showarticletitle{RISC5: Implementing the RISC-V ISA in gem5}. In
  \bibinfo{booktitle}{\emph{Proceedings of Computer Architecture Research with
  RISC-V}}.
\newblock


\bibitem[\protect\citeauthoryear{{Rosenfeld}, {Cooper-Balis}, and
  {Jacob}}{{Rosenfeld} et~al\mbox{.}}{2011}]%
        {dramsim2}
\bibfield{author}{\bibinfo{person}{P. {Rosenfeld}}, \bibinfo{person}{E.
  {Cooper-Balis}}, {and} \bibinfo{person}{B. {Jacob}}.}
  \bibinfo{year}{2011}\natexlab{}.
\newblock \showarticletitle{DRAMSim2: A Cycle Accurate Memory System
  Simulator}.
\newblock \bibinfo{journal}{\emph{IEEE Computer Architecture Letters}}
  \bibinfo{volume}{10}, \bibinfo{number}{1} (\bibinfo{year}{2011}),
  \bibinfo{pages}{16--19}.
\newblock


\bibitem[\protect\citeauthoryear{{Sandberg}, {Nikoleris}, {Carlson},
  {Hagersten}, {Kaxiras}, and {Black-Schaffer}}{{Sandberg}
  et~al\mbox{.}}{2015}]%
        {full-speed-ahead}
\bibfield{author}{\bibinfo{person}{A. {Sandberg}}, \bibinfo{person}{N.
  {Nikoleris}}, \bibinfo{person}{T.~E. {Carlson}}, \bibinfo{person}{E.
  {Hagersten}}, \bibinfo{person}{S. {Kaxiras}}, {and} \bibinfo{person}{D.
  {Black-Schaffer}}.} \bibinfo{year}{2015}\natexlab{}.
\newblock \showarticletitle{Full Speed Ahead: Detailed Architectural Simulation
  at Near-Native Speed}. In \bibinfo{booktitle}{\emph{2015 IEEE International
  Symposium on Workload Characterization}}. \bibinfo{pages}{183--192}.
\newblock


\bibitem[\protect\citeauthoryear{Sardashti, Arelakis, Stenstr{\"o}m, and
  Wood}{Sardashti et~al\mbox{.}}{2015}]%
        {sardashti2015primer}
\bibfield{author}{\bibinfo{person}{Somayeh Sardashti}, \bibinfo{person}{Angelos
  Arelakis}, \bibinfo{person}{Per Stenstr{\"o}m}, {and}
  \bibinfo{person}{David~A Wood}.} \bibinfo{year}{2015}\natexlab{}.
\newblock \showarticletitle{A primer on compression in the memory hierarchy}.
\newblock \bibinfo{journal}{\emph{Synthesis Lectures on Computer Architecture}}
  \bibinfo{volume}{10}, \bibinfo{number}{5} (\bibinfo{year}{2015}),
  \bibinfo{pages}{1--86}.
\newblock


\bibitem[\protect\citeauthoryear{Shao, Xi, Srinivasan, Wei, and Brooks}{Shao
  et~al\mbox{.}}{2016}]%
        {ShaoXSWB16}
\bibfield{author}{\bibinfo{person}{Yakun~Sophia Shao},
  \bibinfo{person}{Sam~Likun Xi}, \bibinfo{person}{Vijayalakshmi Srinivasan},
  \bibinfo{person}{Gu{-}Yeon Wei}, {and} \bibinfo{person}{David~M. Brooks}.}
  \bibinfo{year}{2016}\natexlab{}.
\newblock \showarticletitle{Co-designing accelerators and SoC interfaces using
  gem5-Aladdin}. In \bibinfo{booktitle}{\emph{49th Annual {IEEE/ACM}
  International Symposium on Microarchitecture, {MICRO} 2016, Taipei, Taiwan,
  October 15-19, 2016}}. \bibinfo{publisher}{{IEEE} Computer Society},
  \bibinfo{pages}{48:1--48:12}.
\newblock
\urldef\tempurl%
\url{https://doi.org/10.1109/MICRO.2016.7783751}
\showDOI{\tempurl}


\bibitem[\protect\citeauthoryear{Shingarov}{Shingarov}{2015}]%
        {Shingarov2015-jit}
\bibfield{author}{\bibinfo{person}{Boris Shingarov}.}
  \bibinfo{year}{2015}\natexlab{}.
\newblock \showarticletitle{{Live Introspection of Target-Agnostic JIT in
  Simulation}}. In \bibinfo{booktitle}{\emph{{Proceedings of the International
  Workshop on Smalltalk Technologies}}} \emph{(\bibinfo{series}{IWST})}.
  \bibinfo{pages}{5:1--5:9}.
\newblock
\urldef\tempurl%
\url{https://doi.org/10.1145/2811237.2811295}
\showDOI{\tempurl}


\bibitem[\protect\citeauthoryear{Spiliopoulos, Bagdia, Hansson, Aldworth, and
  Kaxiras}{Spiliopoulos et~al\mbox{.}}{2013}]%
        {SpiliopoulosBHAK13}
\bibfield{author}{\bibinfo{person}{Vasileios Spiliopoulos},
  \bibinfo{person}{Akash Bagdia}, \bibinfo{person}{Andreas Hansson},
  \bibinfo{person}{Peter Aldworth}, {and} \bibinfo{person}{Stefanos Kaxiras}.}
  \bibinfo{year}{2013}\natexlab{}.
\newblock \showarticletitle{Introducing DVFS-Management in a Full-System
  Simulator}. In \bibinfo{booktitle}{\emph{2013 {IEEE} 21st International
  Symposium on Modelling, Analysis and Simulation of Computer and
  Telecommunication Systems, San Francisco, CA, USA, August 14-16, 2013}}.
  \bibinfo{publisher}{{IEEE} Computer Society}, \bibinfo{pages}{535--545}.
\newblock
\urldef\tempurl%
\url{https://doi.org/10.1109/MASCOTS.2013.75}
\showDOI{\tempurl}


\bibitem[\protect\citeauthoryear{Steiner, Jung, Prado, Bykov, and Wehn}{Steiner
  et~al\mbox{.}}{2020}]%
        {stejun_20}
\bibfield{author}{\bibinfo{person}{Lukas Steiner}, \bibinfo{person}{Matthias
  Jung}, \bibinfo{person}{Felipe~S. Prado}, \bibinfo{person}{Kyrill Bykov},
  {and} \bibinfo{person}{Norbert Wehn}.} \bibinfo{year}{2020}\natexlab{}.
\newblock \showarticletitle{{DRAMS}ys4.0: {A} {F}ast and {C}ycle-{A}ccurate
  {S}ystem{C}/{TLM}-{B}ased {DRAM} {S}imulator}. In
  \bibinfo{booktitle}{\emph{International Conference on Embedded Computer
  Systems Architectures Modeling and Simulation (SAMOS)}}.
  \bibinfo{publisher}{Springer}.
\newblock


\bibitem[\protect\citeauthoryear{Ta, Cheng, and Batten}{Ta
  et~al\mbox{.}}{2018}]%
        {risc5-multicore-gem5}
\bibfield{author}{\bibinfo{person}{Tuan Ta}, \bibinfo{person}{Lin Cheng}, {and}
  \bibinfo{person}{Christopher Batten}.} \bibinfo{year}{2018}\natexlab{}.
\newblock \showarticletitle{Simulating Multi-Core RISC-V Systems in gem5}. In
  \bibinfo{booktitle}{\emph{Proceedings of Computer Architecture Research with
  RISC-V}}.
\newblock


\bibitem[\protect\citeauthoryear{Ta, Zhang, Gutierrez, and Beckmann}{Ta
  et~al\mbox{.}}{2019}]%
        {Ta2019gputesting}
\bibfield{author}{\bibinfo{person}{Tuan Ta}, \bibinfo{person}{Xianwei Zhang},
  \bibinfo{person}{Anthony Gutierrez}, {and} \bibinfo{person}{Bradford~M.
  Beckmann}.} \bibinfo{year}{2019}\natexlab{}.
\newblock \showarticletitle{Autonomous Data-Race-Free {GPU} Testing}. In
  \bibinfo{booktitle}{\emph{{IEEE} International Symposium on Workload
  Characterization, {IISWC} 2019, Orlando, FL, USA, November 3-5, 2019}}.
  \bibinfo{publisher}{{IEEE}}, \bibinfo{pages}{81--92}.
\newblock
\urldef\tempurl%
\url{https://doi.org/10.1109/IISWC47752.2019.9042019}
\showDOI{\tempurl}


\bibitem[\protect\citeauthoryear{Tanimoto, Ono, and Inoue}{Tanimoto
  et~al\mbox{.}}{2017}]%
        {tanimoto2017dependence}
\bibfield{author}{\bibinfo{person}{Teruo Tanimoto}, \bibinfo{person}{Takatsugu
  Ono}, {and} \bibinfo{person}{Koji Inoue}.} \bibinfo{year}{2017}\natexlab{}.
\newblock \showarticletitle{Dependence Graph Model for Accurate Critical Path
  Analysis on Out-of-Order Processors}.
\newblock \bibinfo{journal}{\emph{Journal of Information Processing}}
  \bibinfo{volume}{25} (\bibinfo{year}{2017}), \bibinfo{pages}{983--992}.
\newblock


\bibitem[\protect\citeauthoryear{{The HDF Group}}{{The HDF Group}}{2020}]%
        {hdf5}
\bibfield{author}{\bibinfo{person}{{The HDF Group}}.}
  \bibinfo{year}{2020}\natexlab{}.
\newblock \bibinfo{title}{{The HDF5 Library \& File Format}}.
\newblock
  \bibinfo{howpublished}{\url{https://www.hdfgroup.org/solutions/hdf5/}}.
\newblock


\bibitem[\protect\citeauthoryear{Walker, Bischoff, Diestelhorst, Merrett, and
  Al-Hashimi}{Walker et~al\mbox{.}}{2018}]%
        {walker2018hardware}
\bibfield{author}{\bibinfo{person}{Matthew Walker}, \bibinfo{person}{Sascha
  Bischoff}, \bibinfo{person}{Stephan Diestelhorst}, \bibinfo{person}{Geoff
  Merrett}, {and} \bibinfo{person}{Bashir Al-Hashimi}.}
  \bibinfo{year}{2018}\natexlab{}.
\newblock \showarticletitle{Hardware-Validated CPU Performance and Energy
  Modelling}. In \bibinfo{booktitle}{\emph{Performance Analysis of Systems and
  Software (ISPASS), 2018 IEEE International Symposium on}}. IEEE,
  \bibinfo{pages}{44--53}.
\newblock


\bibitem[\protect\citeauthoryear{Wang, Ganesh, Tuaycharoen, Baynes, Jaleel, and
  Jacob}{Wang et~al\mbox{.}}{2005}]%
        {wang_05}
\bibfield{author}{\bibinfo{person}{David Wang}, \bibinfo{person}{Brinda
  Ganesh}, \bibinfo{person}{Nuengwong Tuaycharoen}, \bibinfo{person}{Kathleen
  Baynes}, \bibinfo{person}{Aamer Jaleel}, {and} \bibinfo{person}{Bruce
  Jacob}.} \bibinfo{year}{2005}\natexlab{}.
\newblock \showarticletitle{DRAMsim: A Memory System Simulator}.
\newblock \bibinfo{journal}{\emph{SIGARCH Compututer Architecture News}}
  \bibinfo{volume}{33}, \bibinfo{number}{4} (\bibinfo{date}{Nov.}
  \bibinfo{year}{2005}), \bibinfo{pages}{100--107}.
\newblock
\showISSN{0163-5964}
\urldef\tempurl%
\url{https://doi.org/10.1145/1105734.1105748}
\showDOI{\tempurl}


\bibitem[\protect\citeauthoryear{Waterman, Lee, Patterson, and
  Asanovi{\'c}}{Waterman et~al\mbox{.}}{2011}]%
        {Waterman2011riscv}
\bibfield{author}{\bibinfo{person}{Andrew Waterman}, \bibinfo{person}{Yunsup
  Lee}, \bibinfo{person}{David~A. Patterson}, {and} \bibinfo{person}{Krste
  Asanovi{\'c}}.} \bibinfo{year}{2011}\natexlab{}.
\newblock \bibinfo{booktitle}{\emph{The RISC-V Instruction Set Manual, Volume
  I: Base User-Level ISA}}.
\newblock \bibinfo{type}{{T}echnical {R}eport} UCB/EECS-2011-62.
  \bibinfo{institution}{EECS Department, University of California, Berkeley}.
\newblock
\urldef\tempurl%
\url{http://www2.eecs.berkeley.edu/Pubs/TechRpts/2011/EECS-2011-62.html}
\showURL{%
\tempurl}


\end{thebibliography}

\end{document}